\pdfoutput=1
\documentclass[12pt]{article}
\usepackage{hyperref}
\usepackage{amsmath}
\usepackage{graphicx,psfrag,epsf}
\usepackage{enumerate}
\usepackage[round]{natbib}
\usepackage{url} 
\usepackage{setspace}
\usepackage{authblk}
\usepackage{amsthm}
\usepackage[mathscr]{euscript}
\usepackage{mathtools}
\usepackage{bm}
\usepackage{multirow}
\usepackage{multicol}
\usepackage{diagbox}
\usepackage{placeins}
\usepackage{subfigure}
\usepackage{bigints}
\usepackage{doi}
\usepackage{hyperref}
\usepackage[margin=1in]{geometry} 
\usepackage{makecell}
  
\newcommand\independent{\protect\mathpalette{\protect\independenT}{\perp}}
\def\independenT#1#2{\mathrel{\rlap{$#1#2$}\mkern2mu{#1#2}}}
\theoremstyle{definition}
\newtheorem{definition}{Definition}
\begin{document}
\date{}

  \title{\bf Leveraging Random Assignment to Impute Missing Covariates in Causal Studies\\}
  \author[a]{Gauri Kamat\thanks{Corresponding author. Email: gaurik@brown.edu}}
\author[b]{Jerome P. Reiter\thanks{Research supported by NSF award SES-11-31897}}
\affil[a]{\textit{Department of Biostatistics, Brown University}}
\affil[b]{\textit{Department of Statistical Science, Duke University}}
  
\maketitle

\bigskip

\begin{abstract}
Baseline covariates in randomized experiments are often used in the estimation of treatment effects, for example, when estimating treatment effects within covariate-defined subgroups. In practice, however, covariate values may be missing for some data subjects. To handle missing values, analysts can use imputation methods to create completed datasets, from which they can estimate treatment effects. Common imputation methods include mean imputation, single imputation via regression, and multiple imputation. For each of these methods, we investigate the benefits of leveraging randomized treatment assignment in the imputation routines, that is, making use of the fact that the true covariate distributions are the same across treatment arms. We do so using simulation studies that compare the quality of inferences when we respect or disregard the randomization. We consider this question for imputation routines implemented using covariates only, and imputation routines implemented using the outcome variable. In either case, accounting for randomization offers only small gains in accuracy for our simulation scenarios. Our results also shed light on the  performances of these different procedures for imputing missing covariates in randomized experiments when one seeks to estimate heterogeneous treatment effects.
\end{abstract}


\section{Introduction} \label{sec:1}
Randomized experiments are widely considered to be the gold standard for causal inference. Their appeal, in large part, is attributed to randomized treatment assignment, which ensures baseline comparability of all treatment groups. Alternatively stated, randomization balances all covariates on average \citep{rubin2008}, facilitating simple causal comparisons. Randomized studies can be broadly separated into two stages, namely (1) the stage where the experiment is planned, covariates measured, and treatments assigned, which \citet{rubin2007} calls the ``design stage'', and (2) the stage where the outcome variable is measured and compared across treatment groups, which we henceforth call the ``outcome stage''. Since the design stage is executed without any outcome data in view, results do not systematically favor particular treatment groups \citep{rubin2008}. \par
As with all studies, the data from randomized experiments can suffer from missing values, both in the outcomes and the covariates. While much research focuses on missing outcomes \citep[e.g.,][]{frangakisandrubin1999,chenetal2009,imai2009}, we focus on missing covariates here, assuming for didactic reasons that no outcome values are missing. In this situation, analysts who estimate average treatment effects with simple comparisons of outcome means can disregard the missingness in the covariates. However, analysts generally cannot do so when they estimate subgroup treatment effects or use regression adjustment. In such cases, a complete case analysis (CCA) sacrifices information and can result in distorted estimates. Moreover, CCA violates the intention to treat (ITT) principle in randomized studies \citep{whiteandthompson2005}. \par
An alternative is to impute values for the missing observations. For example, in mean imputation, the analyst replaces each missing value in the covariate $X$ with the mean of the observed values of $X$.  In stochastic regression imputation, the analyst replaces each missing value in $X$ with a draw from a  regression model relating $X$ to other covariates and possibly the treatment or the outcome. These single imputation techniques for missing covariates have been described in detail by \citet{schempersmith1990} and \citet{whiteandthompson2005}.
In multiple imputation (MI) \citep{rubin1987, rubin1996,littlerubin2002}, the analyst creates multiple completed datasets by filling in missing values with draws from predictive distributions estimated with the observed data. The analyst performs complete-data causal inference on each completed dataset, and combines the results using the inferential methods of \citet{rubin1987}. \par
When implementing any of these techniques in randomized experiments, analysts can utilize the randomization in the imputation routine. In MI, for example, they can require the imputed data to be samples from the same covariate distribution, regardless of treatment assignment. This may allow analysts to estimate parameters of the MI model more accurately than estimating separate imputation models in each treatment arm. In turn, this could reduce variances in the imputations and increase accuracy in estimating causal effects from the completed datasets. As another example, in mean imputation, the analyst could pool the observed data across treatment arms to estimate the covariate means,  as opposed to estimating covariate means separately in each treatment arm. \par
In this article, we investigate the magnitude of such improvements in moderately-sized randomized experiments for MI, regression imputation, and mean imputation. For all methods, we consider imputation routines that adhere to the tenets of \citet{rubin2007,rubin2008} by keeping the design and outcome stages separate; that is, we do not include the outcome in the imputation routine. We also consider methods that use the outcome in the imputation routine, as recommended by other authors \citep[e.g.,][]{vach1994,moons2006}. We assess the benefits of respecting or not respecting the randomization, as well as the benefits of using or not using the outcome variable, via simulation studies based on  an ignorable and a non-ignorable missing data scenario. We also use the simulation results to discuss the performances of the single imputation and the MI approaches more generally. \par
Our results contribute to previous work on handling missing covariates in randomized experiments \citep[e.g., ][]{whiteandthompson2005,groenwoldetal2012,jolani2017,sullivanetal18,kayembeetal2020}. In particular, as part of broader simulation studies, \citet{whiteandthompson2005} compare properties of adjusted treatment effects when performing mean imputation within and across treatment arms. \citet{sullivanetal18} compare properties of causal estimates when performing MI with a treatment indicator included in the imputation model, and separately within treatment arms. These works use methods and run simulations for a single covariate, whereas we use more than one covariate. They focus on estimating outcome models assuming constant treatment effects, whereas we estimate outcome models where treatment effects are modified by covariate values. Finally, they use simulations with smaller sample sizes than we do.  As a result of these differences, we are able to reach additional conclusions about the benefits of leveraging randomization, and we contribute new insights into the performances of the different imputation methods.\par
As part of the simulations, we develop and illustrate an MI approach for covariate values that ensures non-parametric identification \citep{vansteelandtetal2006,Robins1997}. This approach is particularly suited for covariates that are not missing at random. We describe this approach in Section \ref{sec:3}, as well as in detail in the supplementary materials.\par
The remainder of this article is organized as follows. In Section \ref{sec:2}, we provide background on randomized experiments and on missing data modeling. In Section \ref{sec:3}, we describe how we respect random assignment in the context of MI. In Section \ref{sec:4}, we describe the simulation design and results. In Section \ref{sec:5}, we implement all the methods on data from \citet{foosgilardi2019}. Finally, in Section \ref{sec:6}, we conclude with a discussion.

\section{Background}\label{sec:2}
We begin with notation and key assumptions in randomized experiments. Here and throughout, we assume a parallel design with an active treatment and a control.
\subsection{Randomized experiments}
We consider a randomized experiment with $n$ units. For $i=1, \dots, n$, let $T_i = 1$ when unit $i$ is assigned to the treatment, and $T_i = 0$ when unit $i$ is assigned to the control. For $i=1, \dots, n$, let $X_{ij}$ denote the value of covariate $j$ measured for unit $i$, and let $X_i = (X_{i1}, \dots, X_{ip})$ represent measurements on $p$ covariates of interest. Let $Y_i(1)$ and $Y_i(0)$ be the potential outcomes associated with the treatment and the control, respectively. We assume that all $n$ units' potential outcomes, $(Y_i(1), Y_i(0))$, are independent. For unit $i$, let $Y_i$ be the outcome value observed at the end of the experiment. \par
We derive results under the stable unit treatment value assumption (SUTVA) \citep{rubin1978}. Specifically, we assume that there is only one version of the treatment and no interference between units. With SUTVA, we can express $Y_i$ as a deterministic function of $Y_i(1)$, $Y_i(0)$, and $T_i$, namely $Y_i = T_iY_i(1) + (1-T_i)Y_i(0)$. For a completely randomized experiment under SUTVA, the treatment status $T_i$ is independent of the potential outcomes $(Y_i(1), Y_i(0))$ and the set of covariates $X_i$. We write this as $T_i \independent \{Y_i(1),Y_i(0),X_i\}$, where $i=1, \dots, n$.

\subsection{Missing data modeling}\label{sec:2.2}
When covariate values are non-ignorably missing \citep{rubin1976}, the analyst needs to model the covariates and their missingness indicators. This joint distribution cannot be identified without recourse to generally untestable assumptions \citep{imbensandpizer2000,dingandgeng2014}, also known as identifying restrictions. In this section, we describe the approach that we use to make such assumptions in MI, following extant work \citep[e.g.,][]{lineroanddaniels2018}.\par
For ease of exposition, we momentarily ignore the treatment and the outcome, and consider $p$ generic variables measured on $n$ subjects, i.e., $X_i= (X_{i1}, \dots, X_{ip})$ where $i=1,\dots,n$. We let $D_i=(D_{i1}, \dots, D_{ip})$ denote their missingness indicators, such that $D_{ij} = 1$ when $X_{ij}$ is missing, and $D_{ij}=0$ otherwise, for $i = 1, \dots, n$ and $j = 1, \dots, p$.  In what follows, we forgo the subscript $i$ indexing subjects  to simplify notation.  For example, $X=(X_1, \dots, X_p)$ represents a generic vector of $p$ variables, and $D=(D_1, \dots, D_p)$ represents a generic missingness pattern, i.e., the vector of values for the $p$ missingness indicators.   

Based on  $D$, we can divide $X$ into $X_{obs}$, the observed part of $X$, and $X_{mis}$, the missing part of $X$. The full data distribution, $f(X_{mis},X_{obs},D)$, can be factored into the product of the observed data distribution $f(X_{obs},D)$, and the extrapolation distribution $f(X_{mis}|X_{obs},D)$ \citep{danielsandhogan2008}. Here, $f(X_{obs},D)$ is identifiable from the observed data, whereas $f(X_{mis}|X_{obs},D)$ is not. Thus, we need to construct $f(X_{mis}|X_{obs},D)$, and consequentially $f(X_{mis},X_{obs},D)$, by imposing identifying restrictions. 
One reasonable desideratum for specifying $f(X_{mis},X_{obs},D)$, when feasible, is to ensure non-parametric identification \citep{vansteelandtetal2006}, defined below. 
\begin{definition}[Non-parametric Identification]
Let $\mathscr{F}$ denote the class of observed data distributions, and let  $\mathscr{F_A}$ denote a family of full data distributions under restriction(s) $\mathscr{A}$. $\mathscr{F_A}$ is said to be non-parametric identified if given \(f \in \) $\mathscr{F}$, there exists a unique \(f_A \in \) $\mathscr{F_A}$ that marginalizes to \(f\). 
\end{definition}
Several restrictions proposed in the literature fulfill this desideratum \citep[see][and references therein]{lineroanddaniels2018}. In this article, we confine our attention to two, namely the itemwise conditionally independent non-response (ICIN) assumption \citep{sadinlereiter2017} and the missing at random (MAR) assumption \citep{gilletal1997}.
\begin{definition}[Itemwise Conditionally Independent Non-response (ICIN)]
$(X_1, \dots, X_p)$ are missing according to the itemwise conditionally independent non-response assumption when $X_j \independent D_j | X_{-j}, D_{-j}$, for $j=1, \dots, p$. Here, $X_{-j} = (X_{1}, \dots, X_{j-1}, X_{j+1}, \dots, X_{p})$, and $D_{-j}$ is defined likewise. 
\end{definition} 
\noindent In other words, with ICIN we presume that controlling for all other variables and their missingness indicators, the missingness of  $X_j$ does not predict its value and vice versa.
\begin{definition}[Missing at Random (MAR)]
If $(X_1, \dots, X_p)$ are missing at random, then for each missingness pattern $D$, $f(D|X_{mis},X_{obs}) = f(D|X_{obs})$. This can be re-expressed as $f(X_{mis}|X_{obs},D) = f(X_{mis}|X_{obs})$.
\end{definition} 
\noindent With MAR, for every $D$, we assume that the missingness is independent of the missing data conditional on the observed data. In the supplementary materials, we illustrate how ICIN and MAR result in identifiable joint distributions for $p=2$ variables. \par
For MI, we fill in missing values $M>1$ times, by drawing independently from the extrapolation distribution $f(X_{mis}|X_{obs},D)$. This creates $M$ completed data sets, each of which is analyzed separately. Point and variance estimates are combined using the rules in \citet{rubin1987}.

\section{MI methods for handling missing covariates}\label{sec:3}
For purposes of illustration, we consider a randomized experiment with two binary covariates, $(X_1,X_2) \in \{0,1\}^2$, having associated missingness indicators $(D_1, D_2)$ $\in \{0,1\}^2$. The treatment status $T \in \{0,1\}$, and the  binary outcome $Y \in \{0,1\}$, are fully observed. For estimating treatment effects, we focus on analysis models with effect modification.  In particular, we assume the logistic regression, 
\begin{align}
\text{logit }(Pr(Y = 1|X_{1},X_{2},T)) = \beta_0 + \beta_1X_{1} + \beta_2X_{2} + \beta_tT + \beta_{tx_2}TX_{2}
\end{align}
is of interest. The parameters of importance are $\beta_t$ and $\beta_{tx_2}$.

\subsection{Design stage modeling} \label{sec3.1}
Covariates are design stage quantities. To follow the principle that the design stage should remain free of any influence from the outcome, the imputation model for $(X_1,X_2)$ should only include quantities observable at the time of treatment assignment. Hence, in our example, the data used for imputation modeling will be  $(X_1,X_2,D_1,D_2,T)$. This can be expressed as a $2^5$ contingency table.\par
Covariates should be measured before randomizing units to treatment or control. When this happens, it is generally the case that the reasons for missingness in the covariates do not depend on treatment group membership. As a result, it is reasonable to regard $(D_1, D_2)$ as pretreatment variables, implying that $(D_1, D_2)\independent  T$. Since we have $(X_1, X_2)\independent  T$, we can collapse the $2^5$ table over $T$, forming a $2^4$ marginal table for $(X_1,X_2,D_1,D_2)$. Thus, in the design stage, respecting randomization is equivalent to treating $(X_1,X_2,D_1,D_2)$ as full data. MI then proceeds by identifying $f(X_1,X_2,D_1,D_2)$, and drawing imputations from $f(X_{mis}|X_{obs},D_1,D_2)$. \par
Under this method, let $\theta$ denote the vector of probabilities offered by the observed data. We have $\theta = (\theta_1, \dots, \theta_9)$, where,
\begin{align}
\theta_1&=Pr(X_1=0,X_2=0,D_1 = 0,D_2 =0), \label{eq:2}\\
\theta_2&=Pr(X_1=0,X_2=1,D_1 = 0,D_2 =0),\label{eq:3}\\
\theta_3&=Pr(X_1=1,X_2=0,D_1 = 0,D_2 =0),\label{eq:4}\\
\theta_4&=Pr(X_1=1,X_2=1,D_1 = 0,D_2 =0),\label{eq:5}\\
\theta_5&=Pr(X_1=1,D_1 = 0,D_2 =1),\label{eq:6}\\
\theta_6&=Pr(X_1=0,D_1 = 0,D_2 =1),\label{eq:7}\\
\theta_7&=Pr(X_2=1,D_1 = 1,D_2 =0),\label{eq:8}\\
\theta_8&=Pr(X_2=0,D_1 = 1,D_2 =0),\label{eq:9}\\
\theta_9&=Pr(D_1 = 1,D_2 =1).\label{eq:10} 
\end{align}
For MI inference, we treat the corresponding counts as a multinomial sample, and place a Dirichlet$(\frac{1}{9}, \dots,\frac{1}{9})$ prior on $\theta$. We further generate an imputation in three steps. We sample a value for $\theta$ from its posterior distribution, which is also Dirichlet. Using this value, we obtain extrapolation distributions for all missingness patterns under the identifying restriction of choice. We derive these distributions for the ICIN and MAR assumptions in the supplementary materials. For each missing value, we then obtain an imputation from the pattern-specific extrapolation distribution. We repeat these steps $M$ independent times, creating $M$ completed datasets.\par
Alternatively, not utilizing randomization means disregarding the independence of $(X_1, X_2)$ and $(D_1, D_2)$ with respect to $T$. We hence use the full $2^5$ contingency table, and identify $f(X_1, X_2, D_1, D_2, T)$. Imputations are generated from $f(X_{mis}|X_{obs}, D_1, D_2, T)$, analogous to separately imputing within the $T=0$ and $T=1$ groups. Under this method, $\theta$ is a vector of eighteen observed probabilities, given by \eqref{eq:2} - \eqref{eq:10} in each treatment arm. Multiple imputation then proceeds in the aforementioned manner, except that we use a Dirichlet$(\frac{1}{18}, \dots, \frac{1}{18})$ prior for $\theta$. \par
Intuitively, accounting for randomization can offer the potential for improved accuracy in estimating treatment effects. By collapsing over treatment groups, we estimate imputation model parameters using the full study sample. In contrast, by imputing separately within treatment groups, we estimate imputation model parameters in each group, using a smaller sample size. This decreased sample size can result in larger parameter uncertainty, which in turn can result in greater variability in the imputations, and hence the MI inferences.\par

\subsection{Outcome stage modeling}
For missing covariates in regression models, it has been generally recommended that the outcome $Y$ be used in MI \citep{rubinschenker1991,vachblettner1991,greenfinkle1995,barnardmeng1999,littlerubin2002,moons2006,sterneetal2009}. In randomized experiments, this has the disadvantage of allowing $Y$ to directly influence the design stage, which can bring the face validity of the final conclusions into question. At the same time, not controlling for $Y$ in imputations can lead to distorted estimates, particularly when regression-adjusted estimators are of interest. \citet{little1992} elucidates this issue: if a partly missing covariate $X$ is highly predictive of $Y$, then $Y$ will carry information about $X$ that may not be captured by other variables in the imputation model. If $X$ is imputed without using $Y$, then the imputed part of $X$ will have no (conditional) association with $Y$. This could falsely attenuate the  overall covariate-outcome association.\par
In our example, an imputation model in the outcome stage uses 
$(X_1,X_2,D_1,D_2,T,Y)$ as data. This forms a contingency table with $2^6$ cells. One way to allow randomization to play a role here is to collapse this $2^6$ table across $T$. However, this makes a strong assumption that $Y \independent T$, which ultimately could underestimate the treatment effect \citep[see][for an illustrative simulation]{luashmead2018}. A more principled way is to factorize the joint distribution $f(X_{mis},X_{obs},D_1,D_2,T,Y)$ to naturally represent the design and outcome stages as
\begin{align}
f(X_{mis},X_{obs},D_1,D_2,T,Y)
& = f(X_{mis},X_{obs},D_1,D_2,T) \text{   }f(Y|X_{mis},X_{obs},D_1,D_2,T). 
\end{align}
Here, $f(X_{mis},X_{obs},D_1,D_2,T)$ represents the joint distribution of the design stage quantities, and $f(Y|X_{mis},X_{obs},D_1,D_2,T)$ is the entire outcome response surface. Under random assignment, we have $(X_{mis},X_{obs}) \independent T$, and assume $(D_1,D_2)\independent T$, so that we have 
\begin{align}
f(X_{mis}|X_{obs},D_1,D_2,T,Y) & \propto f(X_{mis},X_{obs},D_1,D_2,T,Y)\nonumber \\
& \propto f(X_{mis},X_{obs},D_1,D_2) \text{  }f(Y|X_{mis},X_{obs},D_1,D_2,T). \label{eq:1} 
\end{align}

We can generate imputations from \eqref{eq:1} using the data augmentation strategy introduced by \citet{tannerwong1987}. For a given draw of the parameter vector $\gamma = (\theta,\beta)$, where $\theta = (\theta_1,\dots,\theta_9)$ and $\beta = (\beta_0,\beta_1,\beta_2,\beta_t,\beta_{tx_2})$, we generate an imputation using
\begin{align}
f(X_{mis}|X_{obs},D_1,D_2,T,Y,\gamma) \propto f(X_{mis},X_{obs},D_1,D_2,\theta)\text{ } f(Y|X_{obs},X_{mis},D_1,D_2,T,\beta).
\end{align}
Given the imputed dataset, we update $\theta$ and $\beta$ from their full conditional posterior distributions. For $\theta$, this is a Dirichlet distribution. For $\beta$, the conditional posterior is non-standard; we sample from this distribution using the Polya-Gamma latent variable technique outlined in \citet{polsonetal2013}. \par
In practice, we specify  $f(Y|X_{mis},X_{obs},D_1,D_2,T)$ using the outcome model posited for analysis. Often, however, analysis models of interest do not adjust for the missingness indicators, in which case, the conditional independence assumption $Y \independent (D_1,D_2)|X_1,X_2,T$ is implicitly made. \par
If we ignore randomization, we essentially use the treatment as well as the outcome in the imputation model. This amounts to identifying $f(X_1,X_2,D_1,D_2,T,Y)$, and generating imputations from $f(X_{mis}|X_{obs},D_1,D_2,T,Y)$. Under this method, $\theta$ is a vector of thirty-six observed probabilities, given by \eqref{eq:2} - \eqref{eq:10} in each category of $(Y,T)$. We use a Dirichlet$(\frac{1}{36}, \dots, \frac{1}{36})$ prior for $\theta$, and carry out MI as described in Section \ref{sec3.1}.

\section{Simulation study}\label{sec:4}
We now conduct repeated sampling studies to evaluate the performance of the imputation methods. We use twelve simulation settings, comprising all combinations of three missingness scenarios, two identifying restrictions, and two covariate-outcome associations. We begin by describing the data generation process.

\subsection{Data generation}
We simulate the randomized experiment from our running example for $n=1000$ units. Each unit is randomized to the treatment or the control arm with equal probabilities, i.e., $T$ is generated from a Bernoulli distribution with mean 0.5. We generate $(X_1, X_2)$ and $(D_1, D_2)$ as per the MAR and ICIN assumptions, under three missingness scenarios.
\subsubsection{Scenario 1} 
In scenario 1, we regard $(D_1,D_2)$ as pre-treatment variables, and generate them independently of $T$. \par
To simulate an MAR situation under this scenario, we first draw $X_1 \sim \text{Bernoulli }(0.7)$, followed by $X_2|X_1=1 \sim \text{Bernoulli }(0.6)$ and $X_2|X_1=0 \sim \text{Bernoulli }(0.45)$. Subsequently, we delete at random 35\% of the observations from $X_1$, and 40\% of the observations from $X_2$. The true missingness mechanism is thus missing completely at random (MCAR), which is a special case of MAR. \par
To create missingness as per ICIN, we make use of the relationship between the ICIN assumption and hierarchical loglinear models. A discussion of this relationship and the data generation approach that follows is deferred to the supplementary materials. We begin by generating $D_1 \sim \text{Bernoulli }(0.35)$ and $D_2 \sim \text{Bernoulli }(0.40)$. Next, we split the $2^4$ contingency table for $(X_1,X_2,D_1,D_2)$ into four, partial $2^2$ tables, controlling for $(D_1,D_2)$. Let $m_{(x_1,x_2).(d_1,d_2)}$ represent the expected count for $(X_1,X_2) = (x_1,x_2)$ in the partial table with $(D_1,D_2) = (d_1,d_2)$. We simulate $m_{(x_1,x_2).(d_1,d_2)}$ from the loglinear model 
\begin{align}
\text{log } m_{(x_1,x_2).(d_1,d_2)} &=  5 + 0.3x_1 - 0.5x_2 + 0.009d_1 + 0.05d_2 \nonumber \\
&\quad \quad +0.5x_1x_2 + 0.75x_1d_2 + 1x_2d_1 + 0.25d_1d_2.
\end{align}
\noindent Further, we obtain multinomial probabilities $\pi_{(x_1,x_2).(d_1,d_2)}$ using \citep{agresti}
\begin{align}
\pi_{(x_1,x_2).(d_1,d_2)} = \frac{m_{(x_1,x_2).(d_1,d_2)}}{\sum_{x_1}\sum_{x_2}m_{(x_1,x_2).(d_1,d_2)}}. 
\end{align}
For each missingness pattern $(D_1, D_2) = (d_1, d_2)$, we jointly draw $(X_1, X_2)=(x_1, x_2)$ with probability $\pi_{(x_1, x_2).(d_1, d_2)}$, and set $X_j$ to missing wherever $D_j = 1$, where $j = 1, 2$. Similar to the MAR setting, this approach (asymptotically) gives $P(X_1=1)\approx 0.7$, $P(X_2=1|X_1=0)\approx 0.45$, and $P(X_2=1|X_1=1)\approx 0.6$. To illustrate, we display counts for one random draw from the ICIN data generation procedure in Table \ref{tab4.1}.

\begin{table}
\centering
\label{Representative Observed Data under ICIN}
\caption[Observed Data under ICIN]{Counts for the observed data generated as per ICIN using the loglinear model approach, for a fixed random seed. NA represents a missing value.\label{tab4.1}}
\begin{tabular}{|c|c|c|c|}
\cline{1-4}
                                                           & \multirow{2}{*}{\bm{$X_2 = 0$}} & \multirow{2}{*}{\bm{$X_2 = 1$}} & \multirow{2}{*}{\bm{$X_2 = \textbf{NA}$}} \\
                                                           &                                 &                          &                                      \\ \hline
\multicolumn{1}{|c|}{\multirow{2}{*}{\bm{$X_1 = 0$}}}  & \multirow{2}{*}{103}                 & \multirow{2}{*}{55}                 & \multirow{2}{*}{65}                 \\
\multicolumn{1}{|c|}{}                                     &                                 &                                     &                                      \\ \hline
\multicolumn{1}{|c|}{\multirow{2}{*}{\bm{$X_1 = 1$}}}  & \multirow{2}{*}{97}                & \multirow{2}{*}{133}                  & \multirow{2}{*}{197}                 \\
\multicolumn{1}{|c|}{}                                     &                                 &                                     &                                      \\ \hline
\multicolumn{1}{|c|}{\multirow{2}{*}{\bm{$X_1 = \textbf{NA}$}}} & \multirow{2}{*}{68}        & \multirow{2}{*}{155}                & \multirow{2}{*}{127}                 \\
\multicolumn{1}{|c|}{}                                     &                                 &                                     &                                      \\ \hline

\end{tabular}
\end{table}

We generate $Y$ from a Bernoulli distribution with probabilities defined by 
\begin{align}
\text{logit }(Pr(Y= 1|X_{1},X_{2},T)) = \beta_1X_{1} + \beta_2X_{2} + \beta_t T + \beta_{tx_2}TX_{2}.
\end{align}
We set $\beta_t = 0.3$. We vary the strength of the association between $(X_1,X_2)$ and $Y$ in two settings. In the high association setting, we set $\beta_1 = 0.8, \beta_2= 0.9$, and $\beta_{tx_{2}} = 0.5$. In the low association setting, we set $\beta_1= 0.02$, $\beta_2= 0.05$, and $\beta_{tx_{2}} = 0.015$. 

\subsubsection{Scenario 2}
In scenario 2, we regard $(D_1,D_2)$ to be post-treatment variables.  Here, data are generated using the models for $(X_1, X_2, T, Y)$ from scenario 1, except that we draw $D_1$ and $D_2$ from the conditional distributions $D_1|T=1 \sim \text{Bernoulli }(0.35)$, $D_1|T=0 \sim \text{Bernoulli }(0.1)$, and $D_2|T=1 \sim \text{Bernoulli }(0.40)$, $D_2|T=0 \sim \text{Bernoulli }(0.1)$.  Thus, missingness rates differ by treatment arms, but do not depend on the covariate values.

\subsubsection{Scenario 3}
In scenario 3, we allow  $(D_1,D_2)$ to be predictive of $Y$.  We generate $(X_1,X_2)$ and $(D_1,D_2)$ as in scenario 1. We linearly adjust for $(D_1,D_2)$ in the outcome generation model, i.e., we use 
\begin{align}
\text{logit }(Pr(Y= 1|X_{1},X_{2},D_1,D_2,T)) = \beta_1X_{1} + \beta_2X_{2} + \beta_tT + \beta_{tx_2}TX_{2} - 0.6D_{1} -0.4D_{2},
\end{align}
where $(\beta_{1}, \beta_2,  \beta_t, \beta_{tx_2})$ are defined as in scenario 1.

\subsection{Methods}
For MI, we create $M=100$ datasets as per the four approaches described in Section \ref{sec:3}, namely respecting randomization in the design stage (MI-R), not respecting randomization in the design stage (MI-NR), respecting randomization in the outcome stage (MI-RY), and not respecting randomization in the outcome stage (MI-NRY). For each imputed dataset, we obtain point and variance estimates for $\beta_t$ and $\beta_{tx_2}$, and combine them for MI inferences. \par
We also examine the performance of mean imputation, stochastic regression imputation, and complete case analysis (CCA). For mean imputation, we replace the missing values for each covariate by the mean of the observed values of that covariate. Although not typically used for categorical data, mean imputation for dichotomous covariates has been validated in \citet{schempersmith1990} and \citet{sullivanetal18} for estimation of average treatment effects. Specifically, we consider mean imputation across treatment arms (Mean-R), in which we use the covariate means from the combined sample of the treated and control cases; mean imputation within treatment arms (Mean-NR), in which we use the covariate means for the observed treated cases in the treatment arm and the observed control cases in the control arm; and, as a way to use the outcome values, mean imputation within each of the four cells in the cross-classification of the treatment and the binary outcome (Mean-NRY). For regression imputation, we replace missing values for a covariate by draws from the posterior predictive distribution obtained from a logistic regression on the other covariate, where each model is estimated using only the complete cases. In particular, we evaluate regression imputation across treatment arms (Reg-R), in which the model is estimated on the combined sample and independently of the treatment; regression imputation within treatment arms (Reg-NR), in which separate regression models are estimated within the treatment and control arms; and, regression imputation within within each of the four cells in the cross-classification of the treatment and the binary outcome (Reg-NRY). \par
In each scenario, we generate 1000 independently sampled, simulated datasets.  For each method, we compute the Monte Carlo estimate of the absolute bias in the treatment effect estimates, the Monte Carlo estimate of the standard deviation (MC-SD) of the treatment effect estimates, the square root of the average estimated variance (i.e., the estimated standard error, SE), the coverage rate of the 95\% confidence intervals (CI), and the average length of the 95\% CI. 

\subsection{Results}
Here, we present results under the ICIN assumption. Results under the MAR assumption are qualitatively similar and are included in the supplementary materials.

\subsubsection{Scenario 1: High association}\label{sec:4.res1}
\FloatBarrier
\begin{figure}[t]
    \centering
    \subfigure{\includegraphics[width=0.46\textwidth]{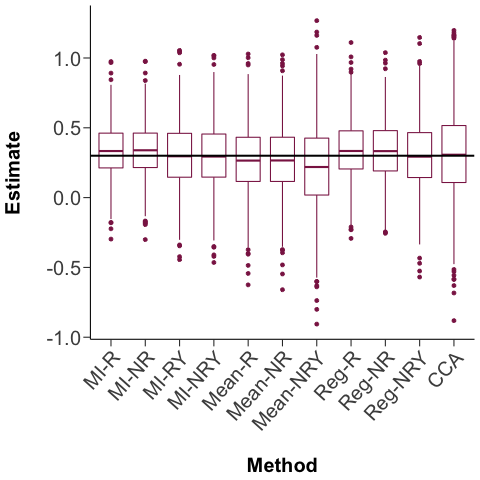}}
    \hspace{0.25in}
    \subfigure{\includegraphics[width=0.46\textwidth]{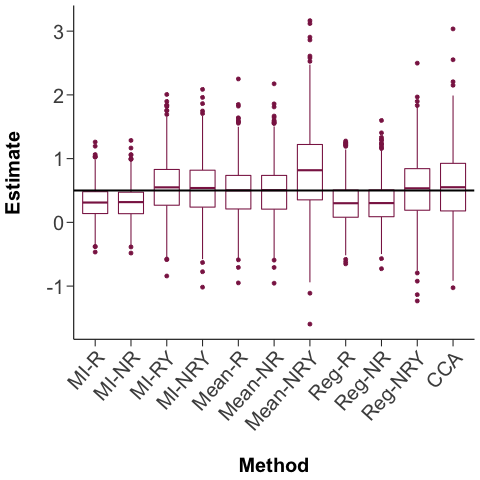}} 
    \vspace{0.4\baselineskip}
    \caption[]{Results for scenario 1 under ICIN for the high association setting. The left panel represents the distribution of $\beta_t$ estimates over 1000 replications, where the true value of $\beta_t = 0.3$. The right panel represents the distribution of $\beta_{tx_2}$ estimates over 1000 replications, where the true value of $\beta_{tx_2} = 0.5$.\label{fig1}}
\end{figure}

\begin{table}[t]
\centering 
\caption[]{Absolute biases, MC standard deviations, estimated standard errors, coverage probabilities, and average CI lengths for $\beta_t$ and $\beta_{tx_2}$ estimates under the high association setting in scenario 1.  True values of $\beta_t=0.3$ and $\beta_{tx_2} = 0.5$. \label{tab:2}}
\resizebox{\textwidth}{!}{
{\LARGE
\begin{tabular}{lcccccccccccc}
\hline
\multicolumn{1}{c}{} & \multicolumn{1}{c}{} & \multicolumn{5}{c}{$\beta_t$} & \multicolumn{1}{c}{} & \multicolumn{5}{c}{$\beta_{tx_2}$} \\ \cline{1-13}  
\multicolumn{2}{l}{\multirow{2}{*}{Method}} &  \multirow{2}{*}{\makecell{Absolute \\ bias}} & \multirow{2}{*}{MC-SD} & \multirow{2}{*}{SE} & \multirow{2}{*}{Coverage}  & \multirow{2}{*}{\makecell{Average CI\\ length}}& \multirow{2}{*}{}     &  \multirow{2}{*}{\makecell{Absolute \\ bias}} & \multirow{2}{*}{MC-SD}      & \multirow{2}{*}{SE} & \multirow{2}{*}{Coverage}  & \multirow{2}{*}{\makecell{Average CI \\ length}} \\ \\
\cline{1-13}  \\
MI-R &&0.039 &0.192  & 0.225 &0.974 &0.881  &&0.192 &0.255 &0.370 &0.978  &1.449 \\
MI-NR &&0.040  &0.192 &0.225  &0.975 &0.884  &&0.194 &0.255  &0.371  &0.983 &1.453 \\
MI-RY &&0.000 &0.232  &0.228 &0.950 &0.890 &&0.021 &0.433 &0.427 &0.946 &1.668  \\ 
MI-NRY &&0.000 &0.233 &0.230 &0.957 &0.903 &&0.022 &0.436 &0.433 &0.947 &1.696 \\
Mean-R &&0.031 &0.241 &0.239 &0.948 &0.938 &&0.018 &0.408 &0.408 &0.949 &1.601\\
Mean-NR &&0.030 &0.249 &0.246 &0.949 &0.940 &&0.017 &0.415 &0.412 &0.947 &1.606\\
Mean-NRY &&0.078 &0.311 &0.234 &0.850 &0.919 &&0.310 &0.616 &0.438 &0.790 &1.715\\
Reg-R &&0.040 &0.210 &0.207 &0.947 &0.813 &&0.199 &0.316 &0.314 &0.902 &1.231 \\
Reg-NR &&0.038 &0.214 &0.210 &0.941 &0.814  &&0.188 &0.322 &0.315 &0.900 &1.234\\
Reg-NRY &&0.001 &0.252 &0.203 &0.884 &0.798 &&0.023 &0.489 &0.335 &0.814 &1.313\\
\multicolumn{1}{l}{CCA}    & \multicolumn{1}{c}{} &  \multicolumn{1}{c}{0.006} & \multicolumn{1}{c}{0.310} & \multicolumn{1}{c}{0.296} & \multicolumn{1}{c}{0.940} &  \multicolumn{1}{c}{1.164} && \multicolumn{1}{c}{0.062}& \multicolumn{1}{c}{0.549} & \multicolumn{1}{c}{0.556} & \multicolumn{1}{c}{0.948} & \multicolumn{1}{c}{2.188} \\\\
\hline \end{tabular}
}
}
\end{table}

Figure \ref{fig1} displays the coefficient estimates for $\beta_t$ and $\beta_{tx_2}$ under scenario 1, when the covariate-outcome association is high. Table \ref{tab:2} displays the absolute biases, MC-SDs, estimated SEs, coverage rates, and average CI lengths for the same scenario. The averages of the $\beta_t$ and $\beta_{tx_2}$ estimates before deletion of data are 0.299 and 0.520, respectively. The MC-SD and 95\% CI coverage rate for the $\beta_t$ estimates before deletion are 0.204 and 0.951, respectively. The MC-SD and 95\% CI coverage rate for the $\beta_{tx_2}$ estimates before deletion are 0.327 and 0.953, respectively. \par
We first consider the MI methods. MI-R and MI-NR produce almost equivalent distributions of point estimates for both coefficients. We also see little difference in the MC-SDs for the two approaches. We ascribe these findings to the relatively large sample size. With 500 units in each treatment arm, we are able to estimate the parameters of the MI model fairly accurately, regardless of whether we combine the data or impute separately in each treatment arm. With binary covariate data, small differences in the precision of the parameters do not substantially alter the posterior predictive distributions for the missing values. The accuracy gain is further diluted, since only a modest fraction of cases are missing. Finally, we note that the analysis model adjusts for both $X_1$ and $X_2$. This can dampen the effect of the accuracy gain even further, as adjustment for covariates is known to increase precision against any residual imbalances that exist in spite of randomization. We see a similarly negligible effect of leveraging randomization when comparing MI-RY and MI-NRY.\par
For $\beta_t$, the outcome stage MI methods produce approximately unbiased estimates (simulated absolute bias $\leq$ 0.001). The simulated absolute biases for their design stage counterparts are higher ($\approx 0.04$). Differences are more pronounced for $\beta_{tx_2}$, with biases for methods MI-R and MI-NR escalating to 0.19 in the negative direction. As $\beta_{tx_2}$ measures how the conditional association between $Y$ and $X_2$ is modified by $T$, this coefficient is attenuated as a result of imputing $X_2$ without using $Y$, explaining the large negative bias. 
\par 
The design stage MI methods exhibit efficiency gains over the outcome stage MI methods, i.e., they have smaller MC-SDs, especially for $\beta_{tx_2}$.  The models that do not control for $Y$ eliminate the uncertainty from estimating coefficients relating the covariates and outcomes in the imputation model. However, this efficiency gain comes at the cost of bias that potentially could be substantial. Interestingly, in these simulations, one could argue that the efficiency gains outweigh the added bias; for example, the simulated mean-squared error is around 0.10 for MI-R and 0.19 for MI-RY.\par
For MI-R and MI-NR, the estimated standard errors for both coefficients are large compared to the corresponding MC-SDs.  This accords with findings that the MI variance estimator can be positively biased \citep{wangrobin1998,robinwang2000,reiterraghunathan2007}, especially when the imputation and analysis models are not congenial \citep{xiemeng2017}. Such large standard errors often lead to wide confidence intervals containing the true parameter values, despite the biases in the point estimates of the coefficients; in fact, both MI-R and MI-NR provide coverage rates over 97\% because of the over-estimation of variances. In contrast, MI-RY and MI-NRY appear not to suffer from over-estimation of variances, and as a result exhibit close to nominal coverage rates.\par


We next evaluate the performance of the mean imputation techniques. Mean-R and Mean-NR perform quite similarly in this scenario, generating nearly identical simulated biases and close to nominal confidence interval coverage rates for both coefficients.   The MC-SDs and estimated SEs under Mean-NR are slightly larger than those under Mean-R, suggesting only minor benefits of leveraging the randomization when using mean imputation for this scenario. 

Mean-NRY is a clear exception among the mean imputation methods.  It generates significantly biased point estimates with the largest MC-SDs, and also has underestimated standard errors. As a result, it has abysmal confidence interval coverage rates.\par 


Turning to the regression imputation approaches, we see that Reg-R and Reg-NR yield similarly biased estimates.  
 The differences in efficiency are modest, with MC-SDs under Reg-NR about 2\% larger than those under Reg-R.  Thus, once again, leveraging randomization seems not to improve inferences meaningfully in this scenario. The biases in the coefficients under Reg-R and Reg-NR are akin to those for MI-R and MI-NR; however, the confidence interval coverage rates are noticeably lower.  This reflects the impact of the bias, which is not compensated for by over-estimation of the variances.

The outcome version of regression imputation, Reg-NRY, offers mixed results.  On the one hand, unlike Reg-R and Reg-NR, it results in coefficient estimates with only small simulated biases. However, the estimated standard errors for $\beta_t$ and $\beta_{tx_2}$ are too small compared to the MC-SDs, resulting in low coverage rates. This is an example of single imputation resulting in under-estimation of variance, emphasized by \citet{rubin1987} as motivation for MI. 

Finally, we consider the performance of CCA. In this scenario, CCA avoids serious biases in $\beta_t$ and $\beta_{tx_2}$,  but it  results in considerably higher MC-SDs than those under competitive imputation methods. Unlike other methods, CCA does not take advantage of the partially observed information. The relatively large variance generates the widest confidence intervals across all methods, although the intervals have close to nominal coverage rates.

Looking across imputation methods, in this scenario, it is evident that Mean-NRY and Reg-NRY are not competitive procedures.  Additionally, if one demands confidence interval coverage rates to be at least near-to-nominal, Reg-R and Reg-NR are dominated by the other imputation procedures. Finally, if one requires low bias in coefficient estimators, Mean-R (or Mean-NR) and MI-RY (or MI-NRY) are most effective.  Arguably, neither method  obviously outperforms the other in these simulations. 

\FloatBarrier
\begin{figure}[t]
    \centering
    \subfigure{\includegraphics[width=0.46\textwidth]{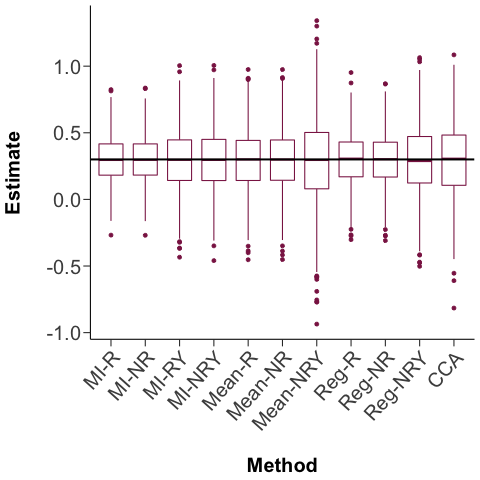}} 
    \hspace{0.25in}
    \subfigure{\includegraphics[width=0.46\textwidth]{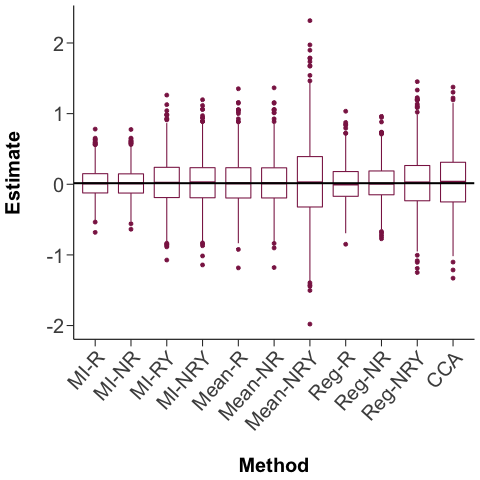}} 
    \vspace{0.4\baselineskip}
    \caption[]{Results for scenario 1 under ICIN for the low association setting. The left panel represents the distribution of $\beta_t$ estimates over 1000 replications, where the true value of $\beta_t = 0.3$. The right panel represents the distribution of $\beta_{tx_2}$ estimates over 1000 replications, where the true value of $\beta_{tx_2} = 0.015$. \label{fig:2}}
\end{figure} 

\subsubsection{Scenario 1: Low association}
Figure \ref{fig:2} displays the  coefficient estimates under scenario 1 when the covariate-outcome association is low. Table \ref{tab:3} provides the corresponding standard errors, coverage rates, and mean CI lengths. The averages of  the $\beta_t$ and $\beta_{tx_2}$ estimates before deletion of data are 0.294 and 0.017 respectively. The MC-SD and 95\% CI coverage rate for the $\beta_t$ estimates before deletion are 0.191 and 0.955, respectively. The MC-SD and 95\% CI coverage rate for the $\beta_{tx_2}$ estimates before deletion are 0.256 and 0.946, respectively. \par
When comparing the design stage methods that leverage random assignment to the corresponding methods that do not, we see at most minor differences in this simulation scenario. MI-R and MI-NR have nearly identical values across all performance metrics. MI-RY is around 1\% to 2\% more efficient than MI-NRY; Mean-R is around 2\% to 4\% more efficient than Mean-NR; and, Reg-R is around 3\% more efficient than Reg-NR. As in the high association scenario, the sample size of 500 in each treatment arm evidently is large enough that the gains in precision in the imputation model parameters do not translate to substantial gains at the treatment effect estimation stage.\par
Looking across methods, we continue to see some of the patterns apparent in the high association setting.  First, for MI-R and MI-NR, the MI variance estimator  continues to be positively biased due to the uncongeniality of the imputation and analysis models.  This results in confidence interval coverage rates around 99\%, even though MI-R and MI-NR have the smallest average confidence interval lengths. Second, MI-RY and MI-NRY have substantially larger MC-SDs than MI-R and MI-NR, although they tend not to have the large positive bias in the MI variance estimator.  We note that the design stage MI methods produce approximately unbiased coefficient estimates. Owing to the low association, the outcome does not carry much information about the partly missing covariates, which reduces the bias due to its exclusion in the covariate imputation model.  Third, Mean-R and Mean-NR continue to perform similarly to MI-RY and MI-NRY.  In this scenario, there is a hint of under-estimation in the variances associated with Mean-R and Mean-NR, which translates to slightly lower than nominal CI coverage rates.  Fourth, Reg-R and Reg-NR continue to offer the smallest average CI lengths.  In this scenario, however, the coverage rates for $\beta_{tx_2}$ are closer to nominal than in Table \ref{tab:2}.  As with the MI methods, ignoring the outcome in the regression imputations does not reduce the quality of the imputations meaningfully when the associations between the covariates and outcome are weak.  Finally, Mean-NRY and Reg-NRY continue to be ineffective as imputation procedures.

\FloatBarrier
\begin{table}[t]
\centering 
\caption[]{Absolute biases, MC standard deviations, estimated standard errors, coverage probabilities, and average confidence lengths for $\beta_t$ and $\beta_{tx_2}$ estimates under the low association setting in scenario 1. True values of $\beta_t=0.3$ and $\beta_{tx_2} = 0.015$. \label{tab:3}}
\resizebox{\textwidth}{!}{
{\LARGE
\begin{tabular}{lcccccccccccc}
\hline
\multicolumn{1}{c}{} & \multicolumn{1}{c}{} & \multicolumn{5}{c}{$\beta_t$} & \multicolumn{1}{c}{} & \multicolumn{5}{c}{$\beta_{tx_2}$} \\ \cline{1-13}  
\multicolumn{2}{l}{\multirow{2}{*}{Method}} &  \multirow{2}{*}{\makecell{Absolute \\ bias}} & \multirow{2}{*}{MC-SD}      & \multirow{2}{*}{SE} & \multirow{2}{*}{Coverage}  & \multirow{2}{*}{\makecell{Average CI\\ length}}& \multirow{2}{*}{}     &  \multirow{2}{*}{\makecell{Absolute \\ bias}} & \multirow{2}{*}{MC-SD} & \multirow{2}{*}{SE} & \multirow{2}{*}{Coverage}  & \multirow{2}{*}{\makecell{Average CI \\ length}} \\ \\
\cline{1-13}  \\
MI-R &&0.002 &0.168 &0.208 &0.989 &0.816  &&0.002 &0.208 &0.302 &0.995 &1.184 \\
MI-NR &&0.002 &0.169 &0.209  &0.988 &0.818 &&0.002 &0.209  &0.303 &0.995 &1.187 \\
MI-RY  &&0.005 &0.222 &0.220 &0.948 &0.858 &&0.012 &0.334 &0.327 &0.948 &1.279 \\ 
MI-NRY &&0.005 &0.225 & 0.224 &0.945 &0.871 &&0.010 &0.340 &0.331 &0.944 &1.309 \\
Mean-R &&0.007 &0.224 &0.219 &0.945 &0.861 &&0.011 &0.340 &0.332 &0.942 &1.303\\
Mean-NR &&0.007 &0.233 &0.225 &0.945 &0.861 &&0.011 &0.345 &0.339 &0.943 &1.303\\
Mean-NRY &&0.014 &0.332&0.220 &0.809 &0.864 &&0.028 &0.569 &0.332 &0.782 &1.307 \\
Reg-R &&0.000 &0.191 &0.189 &0.950 &0.743 &&0.003 &0.268 &0.257 &0.943 &1.009 \\
Reg-NR &&0.003 &0.197 &0.194 &0.946 &0.745 &&0.003 &0.276 &0.267 &0.945 &1.012\\
Reg-NRY &&0.007 &0.251 &0.190 &0.866 &0.744 &&0.012 &0.399 &0.258 &0.797 &1.013\\
\multicolumn{1}{l}{CCA}    & \multicolumn{1}{c}{} &  \multicolumn{1}{c}{0.001}  &\multicolumn{1}{c}{0.286} & \multicolumn{1}{c}{0.279} & \multicolumn{1}{c}{0.947} & \multicolumn{1}{c}{1.094} &&  \multicolumn{1}{c}{0.019} &  \multicolumn{1}{c}{0.419} & \multicolumn{1}{c}{0.416} & \multicolumn{1}{c}{0.940} & \multicolumn{1}{c}{1.631} \\\\ 
\hline \end{tabular}
}
}
\end{table}


\subsubsection{Scenarios 2 and 3}
Under scenarios 2 and 3, conclusions about the missing data methods do not fundamentally change. At $n=1000$, MI methods respecting and not respecting randomization produce similar point estimates, standard errors, and coverage rates. When the association between the covariates and the outcome is high, the design stage MI methods continue to be biased---with biases that are more marked than before---with lower standard errors, while the outcome stage MI methods have low bias with higher standard errors. When this association is low, all MI methods produce comparable point estimates. The single imputation methods and CCA continue to perform as in scenario 1. We present the related graphical and tabular displays in the supplementary materials.

\section{Application}\label{sec:5}
In this section, we present an application of the missing data methods to a randomized experiment analyzed in \citet{foosgilardi2019}, available at \url{https://dataverse.harvard.edu/dataset.xhtml?persistentId=doi:10.7910/DVN/BSIFTF}. The data comprise  $n=612$ women, randomized to receive ($T=1$) or not to receive ($T=0$) an invitation to a career workshop in politics. Approximately two-thirds of the participants receive the treatment.\par
The primary behavioral outcome $Y$ is binary, with $Y=1$ if the participant applies to a political office mentoring program, and $Y=0$ if she does not. Several pre-treatment covariates are measured in the original experiment. We consider two that are highly associated with the outcome, namely an indicator of whether the participant takes an active interest in planning her career ($X_1$), and an indicator of whether the participant wishes to have children in the future ($X_2$). $X_1$ is binary, and exhibits low levels of missingness (2.3\%). $X_2$ contains a ``perhaps/don't know'' category, which we regard as missing data, as in, for example, \citet{rubinetal1995} and \citet{sadinlereiter2017}. This results in 26\% missingness in $X_2$. \par
The distribution of the outcome variable is highly imbalanced, with only 1.5\% of the participants applying to the mentoring program. This leads to analysis models with interaction terms exhibiting perfect prediction issues. We hence focus on estimating the regression-adjusted average treatment effect measured on the log-odds scale, and use the model logit ($\pi$) = $\beta_0 + \beta_1X_1 + \beta_2X_2 + \beta_tT$, where $\pi = Pr(Y=1|X_1,X_2,T)$. The coefficient of interest is $\beta_t$. We note that there is non-compliance in the experiment, since not everyone in the treatment group attended the career workshop. Accordingly, we carry out an ITT analysis. We estimate $\beta_t$ using the same methods as in the simulations in Section \ref{sec:4}. For MI, we use $M=100$ imputations for methods MI-R, MI-NR, and MI-NRY, and iterate to convergence for MI-RY. 
\FloatBarrier
\begin{table}[t]
\centering
\caption[]{Point estimates, standard errors, and 95 \% confidence intervals for the log-odds scale average treatment effect in the randomized experiment from \citet{foosgilardi2019}.\label{tab:4}}
\begin{tabular}{lccc}
\hline
\multirow{2}{*}{Method}    & \multirow{2}{*}{Estimate} & \multirow{2}{*}{SE} & \multirow{2}{*}{95\% CI} \\\\ \hline
                     &                           &                             &                                                   \\

MI-R (ICIN)                  & -0.462                  & 0.678                       &     (-1.790,0.866)   \\

MI-NR (ICIN)                  & -0.462                    & 0.678                       &  (-1.791,0.867)                          \\
MI-RY (ICIN)                 & -0.466                    & 0.669                       &   (-1.794,0.831)                                                            \\
MI-NRY (ICIN)              & -0.469                    & 0.678                      &    (-1.798,0.860)       \\                                              

MI-R (MAR) & -0.464                    & 0.678                      &           (-1.792,0.864)              \\
MI-NR (MAR) & -0.464                    & 0.678                      &          (-1.792,0.865) \\
MI-RY (MAR) & -0.451                   & 0.672                     &    (-1.792,0.827)  \\
MI-NRY (MAR) & -0.467                    & 0.678                     &              (-1.795,0.861)\\
Mean-R &-0.463 &0.677 &(-1.805,0.944)  \\
Mean-NR &-0.463 &0.679 &(-1.805,0.944) \\
Mean-NRY &-0.465 &0.677 &(-1.806,0.943) \\
Reg-R & -0.464 &0.678 & (-1.806,0.944) \\
Reg-NR & -0.468 &0.679 & (-1.810,0.940) \\
Reg-NRY &-0.462 &0.678 &(-1.804,0.945)\\
 CCA                   & -0.692                    & 0.715                       &  (-2.090,0.709)   \\\\ \hline                 
\end{tabular}
\end{table}
Table \ref{tab:4} displays the resultant point estimates, standard errors and 95\% confidence intervals for $\beta_t$. In line with our simulation results, we see that estimates and standard errors are nearly identical for MI-R and MI-NR and also for MI-RY and MI-NRY. The differences between design and outcome stage MI are not remarkable, likely due to the sparsity of $Y=1$ cases in the sample and low missingness levels in the covariates. Results also seem fairly insensitive to the choice of the identifying assumption in MI.  Akin to the simulations, mean imputation as well as the regression imputation approaches perform with efficiency comparable to MI-NRY. All in all, for these data, the choice of imputation method seems not to impact results meaningfully.  \par
We note that CC analysis produces higher standard errors and different point estimates in comparison to the imputation methods.  This difference results mainly from the additional variability in CCA that goes along with deleting the partially observed cases.  Because the outcome has low frequency--which leads to relatively large standard errors--removing the observations with missing covariates has a noticeable effect on the estimates.  

\section{Discussion}\label{sec:6}
Our simulations show that, when units have been properly randomized in the design stage and sample sizes are moderately large, results are practically the same whether one respects or ignores randomization in common methods for imputing missing covariates. This is in slight contrast to the findings by \citet{whiteandthompson2005}, who found that leveraging the randomization can improve inferences when using mean imputation. One key difference in the simulations underpinning those and our findings is the functional form of the outcome model. We estimate non-constant treatment effects for a binary outcome, whereas \citet{whiteandthompson2005} estimate homogeneous treatment effects for a continuous outcome. For these functional forms, it is evidently difficult to realize substantial efficiency gains for the sample sizes we consider.  For randomized experiments where at least one treatment arm has a small size, say in the tens, the gains in precision when accounting for randomization can be more substantial than what we see here.\par 


The results produced by the design and outcome stage imputation methods can differ notably, especially when the covariates are highly prognostic of the outcome and when one seeks to estimate heterogeneous treatment effects. For the MI methods, we observe a trade-off between bias and efficiency. Estimates produced by MI-R and MI-NR can have large bias with relatively low variances, whereas the opposite is true for the outcome stage MI methods. When the covariate-outcome association is low, all four MI methods yield estimates with low simulated bias, although MI-R and MI-NR continue to be more efficient. \par
Interestingly, in these scenarios, MI-R and MI-NR offer higher coverage rates and lower average interval length than MI-RY and MI-NRY. Of course, this pattern need not hold in other simulation scenarios. As an extreme illustration of this point, suppose we have the simulation set-up from Section \ref{sec:4.res1} with very large $n$, say in the 100,000s in each treatment arm.  In this case, the biases in MI-R and MI-NR for $\beta_{tx_2}$, which will persist at the levels in Table \ref{tab:2}, will swamp the MI standard error, even with the variance inflation due to uncongeniality.  Confidence interval coverage rates for MI-R and MI-NR will be well below nominal rates. In contrast, MI-RY and MI-NRY will have biases and coverage rates akin to those seen in Table \ref{tab:2}.  All in all, this suggests that, when one is interested in heterogeneous treatment effects, MI using the outcome potentially has some advantages over MI in the design stage. Of course, this assumes that analysts use a reasonable imputation model, which is hard to be certain of in practice. \par
Turning to the single imputation methods, it is evident that mean imputation in the design stage performs well in these simulations, offering close-to-nominal coverage rates with modest biases and reasonable average confidence interval lengths. These results accord with findings in \citet{sullivanetal18}, who showed that mean imputation  generates small biases in simulations with a single binary covariate and constant treatment effect.  We note that, in large samples, even small biases can overwhelm standard errors, leading to below nominal coverage rates.  Finally, one clear finding from the simulations is that the single imputation methods that account for the outcome perform worse than those that do not, suggesting that the former not be used for imputing missing covariates.\par

Our simulation scenarios are not exhaustive, and results may vary in alternative situations. By design, our findings are specific to analysis models that include an interaction between the treatment and a covariate. They may not apply when constant treatment effects are being estimated. For example, \citet{whiteandthompson2005} show that for a single covariate and no interaction effects, mean imputation in the design stage does not exhibit biases. When covariates are much more strongly associated with one another than in our simulation design, properly specified MI models that account for this association could be notably more efficient than mean imputation. \par


In practice, it is common to adjust for continuous, or a mix of continuous and categorical covariates in the causal analysis. The MI method in Section \ref{sec:2.2} can be adapted to such situations. For example, when $X$ is continuous, we can define the distribution $f(X_{obs},D)$ non-parametrically, using kernel density estimators, as in \citet{titteringtonmill1983}. Alternatively, for each $D$, $f(X_{obs},D)$ can be given a parametric form (see, for e.g., \citet{little1993}, who used normal densities). \citet{sadinlereiter2017} discuss these methods in detail by way of illustrations.\par 
It is also possible that the outcome as well as the covariates contain missing values. In the framework of MI under non-parametric identification, we can separate the covariates and the outcome into blocks and place different identifying restrictions within these blocks. One such block-based method has been presented in \citet{sadinlereiter2018}. We also note that non-parametric identification and subsequent imputation entails breaking down a data set by the observed missingness patterns. For this procedure to work well, sufficient numbers of data points per pattern are required.  In practice, this  amounts to having a large sample size, especially when the data have more than two covariates with missing values.  When sample sizes are not adequate, analysts generally have to sacrifice non-parametric identification.

\begin{center}
{\large\bf SUPPLEMENTARY MATERIALS \\}
We provide two supplements as supporting materials for this article.
\end{center}

\bibliographystyle{myapa} 
\bibliography{mainbibliography.bib}

\newcommand{\noop}[1]{}
\begin{thebibliography}{}
\newcommand{\enquote}[1]{``#1''}

\bibitem[Agresti(2012)Agresti]{agresti}
Agresti, A. (2012), \emph{Categorical Data Analysis}, Hoboken, NJ: John Wiley
  \& Sons.

\bibitem[Barnard and Meng(1999)Barnard and Meng]{barnardmeng1999}
Barnard, J. and Meng, X.~L. (1999), \enquote{Application of multiple imputation
  in medical studies: From AIDS to NHANES,} \emph{Statistical Methods in
  Medical Research}, 8, 17--36.

\bibitem[Chen et~al.(2009)Chen, Geng, and Zhou]{chenetal2009}
Chen, H., Geng, Z., and Zhou, X.~H. (2009), \enquote{Identifiability and
  estimation of causal effects in randomized trials with noncompliance and
  completely nonignorable missing data (with discussion),} \emph{Biometrics},
  65(3), 675--682.

\bibitem[Daniels and Hogan(2008)Daniels and Hogan]{danielsandhogan2008}
Daniels, M.~J. and Hogan, J.~W. (2008), \emph{Missing Data in Longitudinal
  Studies: Strategies for Bayesian Modeling and Sensitivity Analysis}, Boca
  Raton, FL: Chapman and Hall/CRC.

\bibitem[Ding and Geng(2014)Ding and Geng]{dingandgeng2014}
Ding, P. and Geng, Z. (2014), \enquote{Identifiability of subgroup causal
  effects in randomized experiments with nonignorable missing covariates,}
  \emph{Statistics in Medicine}, 33, 1121--1133.

\bibitem[Foos and Gilardi(2019)Foos and Gilardi]{foosgilardi2019}
Foos, F. and Gilardi, F. (2019), \enquote{Does exposure to gender role models
  increase women’s political ambition? A field experiment with politicians,}
  \emph{Journal of Experimental Political Science}, pp. 1--10.

\bibitem[Frangakis and Rubin(1999)Frangakis and Rubin]{frangakisandrubin1999}
Frangakis, C.~E. and Rubin, D.~B. (1999), \enquote{Addressing complications of
  intention-to-treat analysis in the combined presence of all-or-none
  treatment-noncompliance and subsequent missing outcomes,} \emph{Biometrika},
  86(2), 365--379.

\bibitem[Gill et~al.(1997)Gill, van~der Laan, and Robins]{gilletal1997}
Gill, R.~D., van~der Laan, M.~J., and Robins, J.~M. (1997), \enquote{Coarsening
  at random: Characterizations, conjectures, counter-examples,} in
  \emph{Proceedings of the First Seattle Symposium in Biostatistics}, eds.
  D.~Y. Lin and T.~R. Fleming, pp. 255--294.

\bibitem[Greenland and Finkle(1995)Greenland and Finkle]{greenfinkle1995}
Greenland, S. and Finkle, W.~D. (1995), \enquote{A critical look at methods for
  handling missing covariates in epidemiologic regression analyses,}
  \emph{American Journal of Epidemiology}, 142, 1255--1264.

\bibitem[Groenwold et~al.(2012)Groenwold, White, Donders, Carpenter, Altman,
  and Moons]{groenwoldetal2012}
Groenwold, R.~H., White, I.~R., Donders, A.~R., Carpenter, J.~R., Altman,
  D.~G., and Moons, K.~G. (2012), \enquote{Missing covariate data in clinical
  research: When and when not to use the missing-indicator method for
  analysis,} \emph{Canadian Medical Association Journal}, 184(11), 1265--1269.

\bibitem[Imai(2009)Imai]{imai2009}
Imai, K. (2009), \enquote{Statistical analysis of randomized experiments with
  non-ignorable missing binary outcomes: An application to a voting
  experiment,} \emph{Journal of the Royal Statistical Society: Series C
  (Applied Statistics)}, 58(1), 83--104.

\bibitem[Imbens and Pizer(2000)Imbens and Pizer]{imbensandpizer2000}
Imbens, G.~W. and Pizer, W.~A. (2000), \enquote{The analysis of randomized
  experiments with missing data,} \emph{Resources for the Future, Discussion
  Paper}, pp. 00--19.

\bibitem[Jolani and Safarkhani(2017)Jolani and Safarkhani]{jolani2017}
Jolani, S. and Safarkhani, M. (2017), \enquote{The effect of partly missing
  covariates on statistical power in randomized controlled trials with
  discrete-time survival endpoints,} \emph{Methodology: European Journal of
  Research Methods for the Behavioral and Social Sciences}, 13(2), 41--60.

\bibitem[Kayembe et~al.(2020)Kayembe, Jolani, Tan, and {Van
  Breukelen}]{kayembeetal2020}
Kayembe, M.~T., Jolani, S., Tan, F. E.~S., and {Van Breukelen}, G. J.~P.
  (2020), \enquote{Imputation of missing covariate in randomized controlled
  trials with a continuous outcome: Scoping review and new results,}
  \emph{Pharmaceutical Statistics}, pp. 1--21.

\bibitem[Linero and Daniels(2018)Linero and Daniels]{lineroanddaniels2018}
Linero, A.~R. and Daniels, M.~J. (2018), \enquote{Bayesian approaches for
  missing not at random outcome data: The role of identifying restrictions,}
  \emph{Statistical Science}, 33, 198--213.

\bibitem[Little(1992)Little]{little1992}
Little, R. J.~A. (1992), \enquote{Regression with missing X's: A review,}
  \emph{Journal of the American Statistical Association}, 87(420), 1227--1237.

\bibitem[Little(1993)Little]{little1993}
Little, R. J.~A. (1993), \enquote{Pattern-mixture models for multivariate
  incomplete data,} \emph{Journal of the American Statistical Association}, 88,
  125--134.

\bibitem[Little and Rubin(2002)Little and Rubin]{littlerubin2002}
Little, R. J.~A. and Rubin, D.~B. (2002), \emph{Statistical Analysis with
  Missing Data}, Hoboken, NJ: John Wiley \& Sons.

\bibitem[Lu and Ashmead(2018)Lu and Ashmead]{luashmead2018}
Lu, B. and Ashmead, R. (2018), \enquote{Propensity score matching analysis for
  causal effects with MNAR covariates,} \emph{Statistica Sinica}, 28(4),
  2005--2025.

\bibitem[Moons et~al.(2006)Moons, Donders, Stijnen, and Harrell]{moons2006}
Moons, K.~G., Donders, R.~A., Stijnen, T., and Harrell, F.~E. (2006),
  \enquote{Using the outcome for imputation of missing predictor values was
  preferred,} \emph{Journal of Clinical Epidemiology}, 59, 1092--1101.

\bibitem[Polson et~al.(2013)Polson, Scott, and Windle]{polsonetal2013}
Polson, N.~G., Scott, J.~G., and Windle, J. (2013), \enquote{Bayesian inference
  for logistic models using polya-gamma latent variables,} \emph{Journal of the
  American Statistical Association}, 108, 1339--1349.

\bibitem[Reiter and Raghunathan(2007)Reiter and
  Raghunathan]{reiterraghunathan2007}
Reiter, J.~P. and Raghunathan, T.~E. (2007), \enquote{The multiple adaptations
  of multiple imputation,} \emph{Journal of the American Statistical
  Association}, 102, 1462--1471.

\bibitem[Robins(1997)Robins]{Robins1997}
Robins, J.~M. (1997), \enquote{Non-response models for the analysis of
  non-monotone non-ignorable missing data,} \emph{Statistics in Medicine},
  16(1), 21--37.

\bibitem[Robins and Wang(2000)Robins and Wang]{robinwang2000}
Robins, J.~M. and Wang, N. (2000), \enquote{Inference for imputation
  estimators,} \emph{Biometrika}, 87, 113--124.

\bibitem[Rubin(1976)Rubin]{rubin1976}
Rubin, D.~B. (1976), \enquote{Inference and missing data (with discussion),}
  \emph{Biometrika}, 63, 581--592.

\bibitem[Rubin(1978)Rubin]{rubin1978}
Rubin, D.~B. (1978), \enquote{Bayesian inference for causal effects: The role
  of randomization,} \emph{Annals of Statistics}, 6(1), 34--58.

\bibitem[Rubin(1987)Rubin]{rubin1987}
Rubin, D.~B. (1987), \emph{Multiple Imputation for Nonresponse in Surveys}, New
  York, NY: John Wiley \& Sons.

\bibitem[Rubin(1996)Rubin]{rubin1996}
Rubin, D.~B. (1996), \enquote{Multiple imputation after 18+ years,}
  \emph{Journal of the American Statistical Association}, 91(434), 473--489.

\bibitem[Rubin(2007)Rubin]{rubin2007}
Rubin, D.~B. (2007), \enquote{The design versus the analysis of observational
  studies for causal effects: Parallels with the design of randomized trials,}
  \emph{Statistics in Medicine}, 26, 20--30.

\bibitem[Rubin(2008)Rubin]{rubin2008}
Rubin, D.~B. (2008), \enquote{For objective causal inference, design trumps
  analysis,} \emph{Annals of Applied Statistics}, 2 (3), 808--840.

\bibitem[Rubin and Schenker(1991)Rubin and Schenker]{rubinschenker1991}
Rubin, D.~B. and Schenker, N. (1991), \enquote{Multiple imputation in
  health-care databases: An overview and some applications,} \emph{Statistics
  in Medicine}, 10, 585--598.

\bibitem[Rubin et~al.(1995)Rubin, Stern, and Vehovar]{rubinetal1995}
Rubin, D.~B., Stern, H., and Vehovar, V. (1995), \enquote{Handling "Don't Know"
  survey responses: The case of the Slovenian Plebiscite.} \emph{Journal of the
  American Statistical Association}, 90(431), 822--828.

\bibitem[Sadinle and Reiter(2017)Sadinle and Reiter]{sadinlereiter2017}
Sadinle, M. and Reiter, J.~P. (2017), \enquote{Itemwise conditionally
  independent nonresponse modeling for incomplete multivariate data,}
  \emph{Biometrika}, 104(1), 207--220.

\bibitem[Sadinle and Reiter(2018)Sadinle and Reiter]{sadinlereiter2018}
Sadinle, M. and Reiter, J.~P. (2018), \enquote{Sequential identification of
  nonignorable missing data mechanisms,} \emph{Statistica Sinica}, 28,
  1741--1759.

\bibitem[Schemper and Smith(1990)Schemper and Smith]{schempersmith1990}
Schemper, M. and Smith, T.~L. (1990), \enquote{Efficient evaluation of
  treatment effects in the presence of missing covariate values,}
  \emph{Statistics in Medicine}, 9, 777--784.

\bibitem[Sterne et~al.(2009)Sterne, White, Carlin, Spratt, Royston, Kenward,
  Wood, and Carpenter]{sterneetal2009}
Sterne, J. A.~C., White, I.~R., Carlin, J.~B., Spratt, M., Royston, P.,
  Kenward, M.~G., Wood, A.~M., and Carpenter, J.~R. (2009), \enquote{Multiple
  imputation for missing data in epidemiological and clinical research:
  Potential and pitfalls,} \emph{BMJ}, 338:b2393.

\bibitem[Sullivan et~al.(2018)Sullivan, White, Salter, Ryan, and
  Lee]{sullivanetal18}
Sullivan, T.~R., White, I.~R., Salter, A.~B., Ryan, P., and Lee, K.~J. (2018),
  \enquote{Should multiple imputation be the method of choice for handling
  missing data in randomized trials?} \emph{Statistical Methods in Medical
  Research}, 27, 2610--2626.

\bibitem[Tanner and Wong(1987)Tanner and Wong]{tannerwong1987}
Tanner, M. and Wong, W. (1987), \enquote{The calculation of posterior
  distributions by data augmentation,} \emph{Journal of the American
  Statistical Association}, 82(398), 528--540.

\bibitem[Titterington and Mill(1983)Titterington and
  Mill]{titteringtonmill1983}
Titterington, D.~M. and Mill, G.~M. (1983), \enquote{Kernel-based density
  estimates from incomplete data,} \emph{Journal of the Royal Statistical
  Society: Series B (Methodological)}, 45(2), 258--266.

\bibitem[Vach(1994)Vach]{vach1994}
Vach, W. (1994), \emph{Logistic Regression with Missing Values in the
  Covariates}, New York, NY: Springer.

\bibitem[Vach and Blettner(1991)Vach and Blettner]{vachblettner1991}
Vach, W. and Blettner, M. (1991), \enquote{Biased estimates of the odds ratio
  in case-control studies due to the use of ad hoc methods of correcting for
  missing values for confounding variable,} \emph{American Journal of
  Epidemiology}, 134, 895--907.

\bibitem[Vansteelandt et~al.(2006)Vansteelandt, Goetghebeur, Kenward, and
  Molenberghs]{vansteelandtetal2006}
Vansteelandt, S.~R., Goetghebeur, E., Kenward, M., and Molenberghs, G. (2006),
  \enquote{Ignorance and uncertainty regions as inferential tools in a
  sensitivity analysis,} \emph{Statistica Sinica}, 16, 953--979.

\bibitem[Wang and Robins(1998)Wang and Robins]{wangrobin1998}
Wang, N. and Robins, J.~M. (1998), \enquote{Large-sample theory for parametric
  multiple imputation procedures,} \emph{Biometrika}, 85(4), 935--948.

\bibitem[White and Thompson(2005)White and Thompson]{whiteandthompson2005}
White, I.~R. and Thompson, S.~G. (2005), \enquote{Adjusting for partially
  missing baseline measurements in randomized trials,} \emph{Statistics in
  Medicine}, 24, 993--1007.

\bibitem[Xie and Meng(2017)Xie and Meng]{xiemeng2017}
Xie, X. and Meng, X.~L. (2017), \enquote{Dissecting multiple imputation from a
  multi-phase inference perspective: What happens when God{`}s, imputer{`}s and
  analyst{`}s models are uncongenial?} \emph{Statistica Sinica}, 27,
  1485--1594.

\end{thebibliography}

\end{document}


\date{}
\if1\blind
{
  \title{\bf Leveraging Random Assignment to Impute Missing Covariates in Causal Studies - Supplement 1\\}
  \author{Gauri Kamat
    and
    Jerome P. Reiter\\}
  \maketitle
} \fi

\if0\blind
{
  \bigskip
  \bigskip
  \bigskip
  \begin{center}
    {\LARGE\bf Title}
\end{center}
  \medskip
} \fi

\section{Introduction} 
In this supplement, we derive the non-parametric identified distributions used in the multiple imputation (MI) approach in the main text. In Section \ref{sec:B} and Section \ref{sec:C}, we obtain the full data and extrapolation distributions under the ICIN and MAR assumptions, respectively. In Section \ref{sec:D}, we provide mathematical justification for the loglinear model used to simulate expected counts under ICIN. Throughout, we consider variables $X = (X_1,X_2) \in \{0,1\}^2$, with missingness indicators $D = (D_1,D_2) \in \{0,1\}^2$.

\section {Identification under the ICIN assumption} \label{sec:B}
Identification under ICIN follows from Theorem 1 in Sadinle and Reiter (2017). We begin by restating a key formula from the theorem. Define 
\begin{align}
\eta_D(X_{obs}) &= \text{log } f(X_{obs},D) - \text{log} \bigintsss_{X_{mis}} \text{exp} \sum_{D^*<D}\eta_{D^*}(X_{obs})\mathbbm{1}(D^*<D) \text{  } dX_{mis}.
\end{align}
Here, $D^*<D$ if $D$ has strictly greater number of missing elements than $D^*$. For each missingness pattern, the $\eta$ are obtained as follows.
\begin{align}
\eta_{00}(X_1,X_2) &= \text{log } f(X_1,X_2,D_1=0,D_2=0), \\[1em]
\eta_{01}(X_1) &= \text{log } f(X_1,D_1=0,D_2=1)  
- \text{log} \bigintsss_{X_{2}} f(X_1,X_2,D_1=0,D_2=0) \text{ } dX_2 \nonumber \\[1em]
&= \text{log }\dfrac{f(X_1,D_1=0,D_2=1)}{f(X_1,D_1=0,D_2=0)},\\[1em]
\eta_{10}(X_2) &= \text{log } f(X_2,D_1=1,D_2=0) - \text{log} \bigintsss_{X_{1}} f(X_1,X_2,D_1=0,D_2=0) \text{ } dX_1 \nonumber \\[1em]
&= \text{log }\dfrac{f(X_2,D_1=1,D_2=0)}{f(X_2,D_1=0,D_2=0)}, \\[1em]
\noindent \eta_{11} &= \text{log } f(D_1=1,D_2=1) \nonumber \\
&\quad \quad \quad \quad 
- \text{log} \bigintsss_{X_{1}}\bigintsss_{X_{2}}\text{exp } \bigg[\eta_{00}(X_1,X_2) + \eta_{01}(X_1) + \eta_{10}(X_2)\bigg] dX_2 dX_1 \nonumber \\[1em]
&= \text{log }\dfrac{f(D_1=1,D_2=1)}{C}, 
\end{align} 
\noindent where,
\begin{equation}
\begin{split}
C &= \bigintsss_{X_{1}}\bigintsss_{X_{2}}\bigg( \dfrac{f(X_1,X_2,D_1=0,D_2=0)\space f(X_1,D_1=0,D_2=1)} {f(X_2,D_1=1,D_2=0)} \\
 &\quad \quad \quad \quad \qquad \qquad \qquad \qquad \times \dfrac{f(X_1,D_1=0,D_2=0)}{f(X_2,D_1=0,D_2=0)}\bigg) \text{  } dX_2 dX_1. \label{eq:A}\\
\end{split}
\end{equation}
\\
\noindent Further, for each $(D_1,D_2)$ we have
\begin{align}
f(X_1,X_2,D_1,D_2) = \text{exp }\mathop{\mathlarger{\mathlarger {\sum}}}_{D^*\leq D}\eta_{D^*}(X_{obs}).
\end{align}
\noindent Hence,
\begin{align}
f(X_1,X_2,D_1=0,D_2=1) &= \text{exp } \bigg[\eta_{00}(X_1,X_2)
+\eta_{01}(X_1)\bigg] \nonumber \\[1em]
&= \dfrac{f(X_1,X_2,D_1=0,D_2=0) f(X_1,D_1=0,D_2=1)}{f(X_1,D_1=0,D_2=0)}, \\[1em]
f(X_1,X_2,D_1=1,D_2=0) &= \text{exp } \bigg[\eta_{00}(X_1,X_2) +\eta_{10}(X_2)\bigg] \nonumber \\[1em]
&= \dfrac{f(X_1,X_2,D_1=0,D_2=0) f(X_2,D_1=1,D_2=0)}{f(X_2,D_1=0,D_2=0)}, \\[1em]
f(X_1,X_2,D_1=1,D_2=1) &= \text{exp } \bigg[\eta_{00}(X_1,X_2) +\eta_{01}(X_1) + \eta_{10}(X_2) + \eta_{11}\bigg] 
\nonumber\\[1em] 
&= \dfrac{f(X_1,X_2,D_1=0,D_2=0) f(X_1,D_1=0,D_2=1)}{f(X_1,D_1=0,D_2=0)}
\nonumber\\[1em] 
& \quad \quad \quad \quad\times \dfrac{f(X_2,D_1=1,D_2=0)f(D_1=1,D_2=1)}{f(X_2,D_1=0,D_2=0) } \times \dfrac{1}{C},
\end{align}
where $C$ is as defined in \eqref{eq:A}.
\noindent It is straightforward to compute the extrapolation distribution for each $(D_1,D_2)$ using
\begin{align}
 f(X_{mis}|X_{obs},D_1,D_2) \propto f(X_{mis},X_{obs},D_1,D_2).   
\end{align}

\section{Identification under the MAR assumption} \label{sec:C}
Under MAR, the following condition holds:
\begin{align}
f(X_{mis}|X_{obs},D) = f(X_{mis}|X_{obs}) = \displaystyle\sum_{D^{*}\in \mathcal{D}}f(X_{mis}|X_{obs},D^{*})f(D^{*}|X_{obs}), \label{eq:B}
\end{align}
where $\mathcal{D}$ is the set of all missingness patterns that have been realized in the data. In other words, the extrapolation distribution for a particular missingness pattern is obtained by marginalizing across all observed missingness patterns. We note here that an MCAR distribution can also be obtained using \eqref{eq:B}, simply by replacing $f(D^{*}|X_{obs})$ with $f(D^{*})$.\par
When $p=2$, we have $\mathcal{D} = \{(0,0),(0,1),(1,0),(1,1)\}$. For purposes of illustration, we assume that all patterns in $\mathcal{D}$ have been realized. For $(D_1,D_2)$ = $(0,1)$, we have
\begin{align}
f(X_2|X_1,0,1) &= f(X_2|X_1,0,0)f(0,0|X_1)+f(X_2|X_1,0,1)f(0,1|X_1) \nonumber \\ 
&\qquad +  f(X_2|X_1,1,0)f(1,0|X_1) + f(X_2|X_1,1,1)f(1,1|X_1).
\end{align}
\noindent For each pattern in the RHS above, we equate the appropriate extrapolation distribution to the corresponding distribution under $(D_1,D_2)=(0,0)$. That is, for $D^* \in \mathcal{D}$, we  equate $f(X_{mis}|X_{obs},D^*) = f(X_{mis}|X_{obs},0,0)$. Thus, we have
\begin{align}
f(X_2|X_1,0,1) &= f(X_2|X_1,0,0), \\
f(X_2|X_1,1,0) &\propto f(X_1,X_2,1,0)\nonumber \\
&\propto f(X_1|X_2,1,0)f(X_2,1,0)= f(X_1|X_2,0,0)f(X_2,1,0),\\
f(X_2|X_1,1,1) &\propto f(X_1,X_2,1,1) \nonumber \\
&\propto f(X_1,X_2|1,1)f(1,1)= f(X_1,X_2|0,0)f(1,1).
\end{align}
\noindent We similarly identify the extrapolation distribution for $(D_1,D_2) = (1,0)$.
We have
\begin{align}
f(X_1|X_2,1,0) &= f(X_1|X_2,0,0)f(0,0|X_2)+f(X_1|X_2,0,1)f(0,1|X_2) \nonumber \\ 
&\qquad +  f(X_1|X_2,1,0)f(1,0|X_2) + f(X_1|X_2,1,1)f(1,1|X_2), 
\end{align}
\noindent where,
\begin{align}
f(X_1|X_2,1,0) &= f(X_1|X_2,0,0),\\
f(X_1|X_2,0,1) &\propto f(X_1,X_2,0,1) \nonumber \\
&\propto f(X_2|X_1,0,1)f(X_1,0,1)= f(X_2|X_1,0,0)f(X_1,0,1),\\
f(X_1|X_2,1,1) &\propto f(X_1,X_2,1,1) \nonumber \\
&\propto f(X_1,X_2|1,1)f(1,1)= f(X_1,X_2|0,0)f(1,1).
\end{align}
\noindent For $(D_1,D_2) = (1,1)$, we have
\begin{align}
f(X_1,X_2|1,1) &= f(X_1,X_2|0,0)f(0,0)+f(X_1,X_2|0,1)f(0,1) \nonumber\\
\qquad&+  f(X_1,X_2|1,0)f(1,0) + f(X_1,X_2|1,1)f(1,1),
\end{align}
\noindent where,
\begin{align}
f(X_1,X_2|1,1) &= f(X_1,X_2|0,0),\\
f(X_1,X_2|0,1) &\propto f(X_1,X_2,0,1)\nonumber \\
&\propto f(X_2|X_1,0,1)f(X_1,0,1)= f(X_2|X_1,0,0)f(X_1,0,1),\\
f(X_1,X_2|1,0) &\propto f(X_1,X_2,1,0) \nonumber \\
&\propto f(X_1|X_2,1,0)f(X_2,1,0)= f(X_1|X_2,0,0)f(X_2,1,0).
\end{align}
The full data distribution for each $(D_1,D_2)$ can be obtained using 
\begin{align}
f(X_{mis},X_{obs},D_1,D_2) = f(X_{mis}|X_{obs},D_1,D_2)\text{  }f(X_{obs},D_1,D_2).
\end{align}

\section{Data generation under ICIN} \label{sec:D}
We let $(x_1,x_2)$ and $(d_1,d_2)$ denote the values taken by $(X_1,X_2)$ and $(D_1,D_2)$ respectively. For loglinear models, we use the notation established in Agresti (2012). \par
Following Sadinle and Reiter (2017), we regard each term in $\eta$ (defined in Section \ref{sec:B}) to be a hierarchical loglinear model, where the highest order term is a two-way interaction between the observed component of $(X_1,X_2)$ and the indicator of the missing component. Thus, we have
\begin{align}
\eta_{00}(x_1,x_2) &= \text{log }f(X_1,X_2,D_1 = 0,D_2 = 0) \nonumber \\
&= \lambda_{x_1x_2}^{X_1X_2} + \lambda_{x_1}^{X_1} + \lambda_{x_2}^{X_2} + \lambda, \label{eq:C}\\[1em]
\eta_{01}(x_1) &= \text{log }f(X_1,D_1 = 0,D_2 = 1) - \text{log }f(X_1,D_1 = 0,D_2 = 0)\nonumber\\
&= \lambda_{x_11}^{X_1D_2} + \lambda_{x_1}^{X_1} + \lambda_{1}^{D_2} + \lambda - \bigg[\lambda_{x_1}^{X_1} + \lambda\bigg] \nonumber\\
&=\lambda_{x_11}^{X_1D_2} + \lambda_{1}^{D_2},\label{eq:D}\\[1em]
\eta_{10}(x_2) &= \text{log }f(X_2,D_1 = 1,D_2 = 0) - \text{log }f(X_2,D_1 = 0,D_2 = 0)\nonumber\\
&= \lambda_{x_11}^{X_2D_1} + \lambda_{x_2}^{X_2} + \lambda_{1}^{D_1} + \lambda - \bigg[\lambda_{x_2}^{X_2} + \lambda\bigg] \nonumber\\
&= \lambda_{x_21}^{X_2D_1} + \lambda_{1}^{D_1}, \label{eq:E}\\ \nonumber
\end{align}
\begin{align}
\eta_{11} &= \text{log } f(D_1=1,D_2=1)\nonumber\\
& \quad \quad \quad \quad - \text{log} \bigintsss_{X_{1}}\bigintsss_{X_{2}} \text{exp } \bigg[\eta_{00}(x_1,x_2) + \eta_{01}(x_1) + \eta_{10}(x_2)\bigg] dX_2 dX_1 \nonumber \\ 
&=\lambda_{11}^{D_1D_2}. \label{eq:F}
\end{align}

\noindent We now use \eqref{eq:C} - \eqref{eq:F} to express the joint distribution of $(X_1,X_2,D_1,D_2)$ in terms of hierarchical loglinear models. We have
\begin{align}
\text{log }f(X_1,X_2,D_1 = 0, D_2 =0) &= \eta_{00}(x_1,x_2) \nonumber\\ 
&= \lambda +\lambda_{x_1}^{X_1} + \lambda_{x_2}^{X_2} + \lambda_{x_1x_2}^{X_1X_2},\label{eq:G}\\
\text{log }f(X_1,X_2,D_1=0,D_2=1) &=  \eta_{00}(x_1,x_2) + \eta_{01}(x_1)\nonumber\\
&= \lambda + \lambda_{1}^{D_2}+\lambda_{x_1}^{X_1} + \lambda_{x_2}^{X_2}+\lambda_{x_11}^{X_1D_2} + \lambda_{x_1x_2}^{X_1X_2}, \label{eq:H}\\
\text{log }f(X_1,X_2,D_1=1,D_2=0) &= \eta_{00}(x_1,x_2) + \eta_{10}(x_2)\nonumber\\
&= \lambda + \lambda_{1}^{D_1}+\lambda_{x_1}^{X_1} + \lambda_{x_2}^{X_2}+\lambda_{x_21}^{X_2D_1} + \lambda_{x_1x_2}^{X_1X_2},\label{eq:I}\\
\text{log }f(X_1,X_2,D_1=1,D_2=1) &=  \eta_{00}(x_1,x_2) + \eta_{01}(x_1) + \eta_{10}(x_2) + \eta_{11}\nonumber\\
&=\lambda + \lambda_{1}^{D_1}+ \lambda_{1}^{D_2}+ \lambda_{x_1}^{X_1} + \lambda_{x_2}^{X_2}+\lambda_{x_21}^{X_2D_1} +\lambda_{x_11}^{X_1D_2}\nonumber\\ &\quad \quad+\lambda_{x_1x_2}^{X_1X_2}+\lambda_{11}^{D_1D_2}. \label{eq:J}
\end{align}
\eqref{eq:G} - \eqref{eq:J} can be combined into a single loglinear model as 
\begin{align}
\text{log }f(X_1,X_2,D_1,D_2) &=\lambda +  \lambda_{x_1}^{X_1} + \lambda_{x_2}^{X_2} + \lambda_{d_1}^{D_1}+ \lambda_{d_2}^{D_2} +\lambda_{x_2d_1}^{X_2D_1} +\lambda_{x_1d_2}^{X_1D_2} \nonumber \\ &\quad \quad+\lambda_{x_1x_2}^{X_1X_2}+\lambda_{d_1d_2}^{D_1D_2}. \label{eq:K}
\end{align}


\date{}
\if1\blind
{
  \title{\bf Leveraging Random Assignment to Impute Missing Covariates in Causal Studies - Supplement 2\\}
  \author{Gauri Kamat
    and
    Jerome P. Reiter\\}
  \maketitle
} \fi

\if0\blind
{
  \bigskip
  \bigskip
  \bigskip
  \begin{center}
    {\LARGE\bf Title}
\end{center}
  \medskip
} \fi

\section{Introduction} 
In this supplement, we present additional simulation results. In Section \ref{sec:S2}, we present all results under MAR. In Section \ref{sec:S3}, we display additional results under ICIN that are described in the main text.

\section{Simulation results under the MAR assumption} \label{sec:S2}
\begin{figure}[t] 
    \centering
    \subfigure{\includegraphics[width=0.46\textwidth]{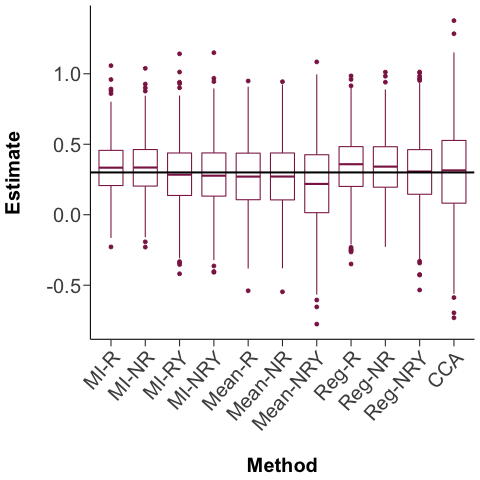}} 
    \hspace{0.25in}
    \subfigure{\includegraphics[width=0.46\textwidth]{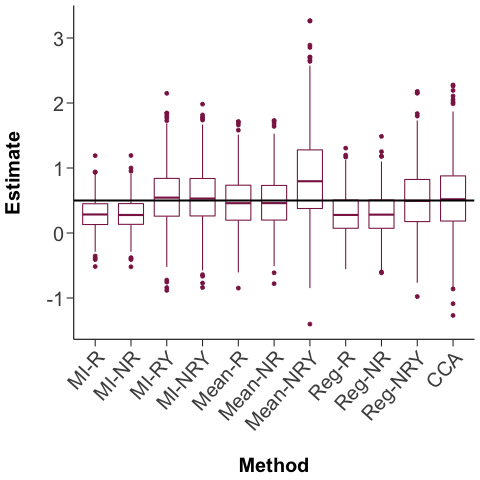}} 
    \vskip\baselineskip
    \subfigure{\includegraphics[width=0.46\textwidth]{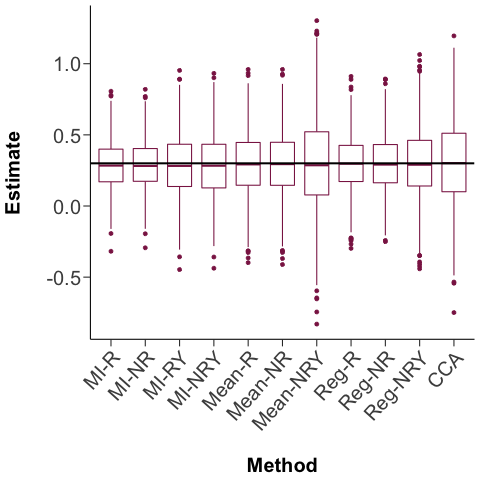}}
    \hspace{0.25in}
    \subfigure{\includegraphics[width=0.46\textwidth]{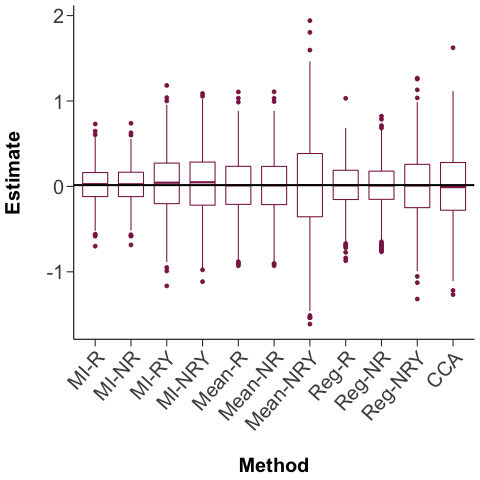}}
    \vspace{0.4\baselineskip}
    \caption[]{Results for scenario 1 under MAR. The top panel represents the distribution of $\beta_t$ (left) and $\beta_{tx_2}$ (right) estimates for the high association setting, where the true values are $\beta_t = 0.3$ and  $\beta_{tx_2} = 0.5$. The bottom panel represents the distribution of $\beta_t$ (left) and $\beta_{tx_2}$ (right) estimates for the low association setting, where the true values are $\beta_t = 0.3$ and  $\beta_{tx_2} = 0.015$. \label{fig:S1}}
\end{figure}

\begin{figure}[t]
    \centering
    \subfigure{\includegraphics[width=0.46\textwidth]{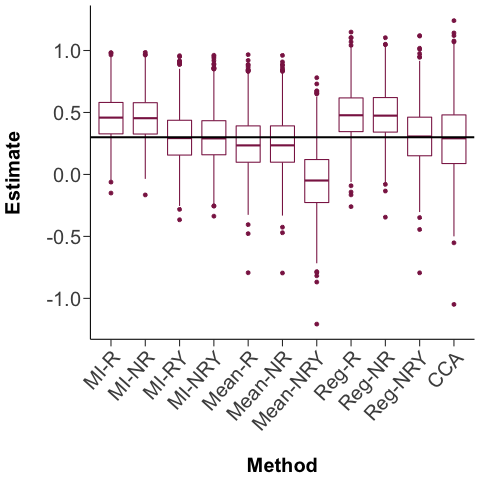}} 
    \hspace{0.25in}
    \subfigure{\includegraphics[width=0.46\textwidth]{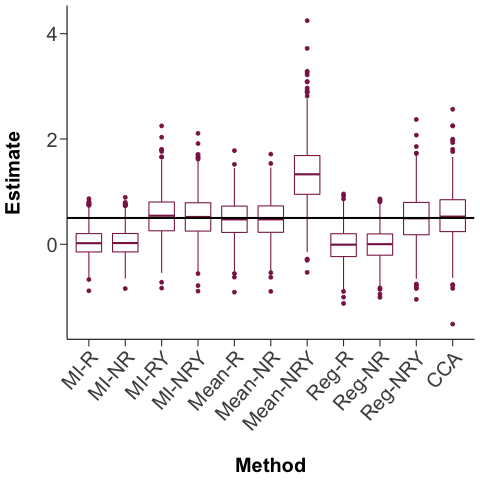}} 
    \vskip\baselineskip
    \subfigure{\includegraphics[width=0.46\textwidth]{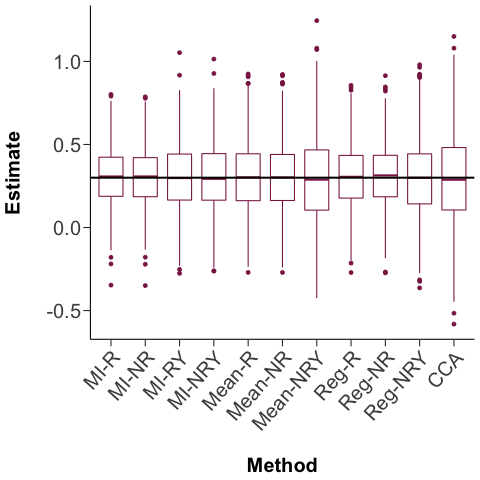}}
    \hspace{0.25in}
    \subfigure{\includegraphics[width=0.46\textwidth]{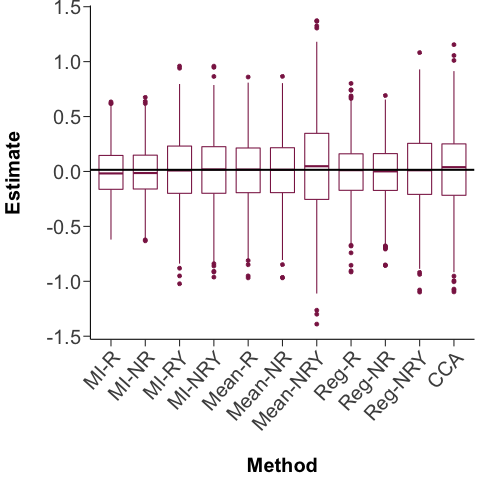}}
    \vspace{0.4\baselineskip}
    \caption[]{Results for scenario 2 under MAR. The top panel represents the distribution of $\beta_t$ (left) and $\beta_{tx_2}$ (right) estimates for the high association setting, where the true values are $\beta_t = 0.3$ and  $\beta_{tx_2} = 0.5$. The bottom panel represents the distribution of $\beta_t$ (left) and $\beta_{tx_2}$ (right) estimates for the low association setting, where the true values are $\beta_t = 0.3$ and  $\beta_{tx_2} = 0.015$.\label{fig:S2}}
\end{figure}

\begin{figure}[t]
    \centering
    \subfigure{\includegraphics[width=0.46\textwidth]{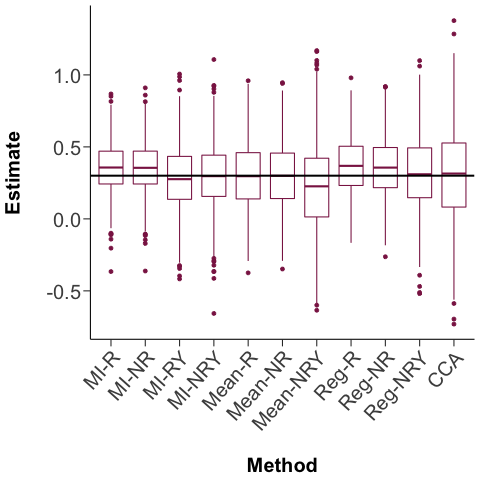}} 
    \hspace{0.25in}
    \subfigure{\includegraphics[width=0.46\textwidth]{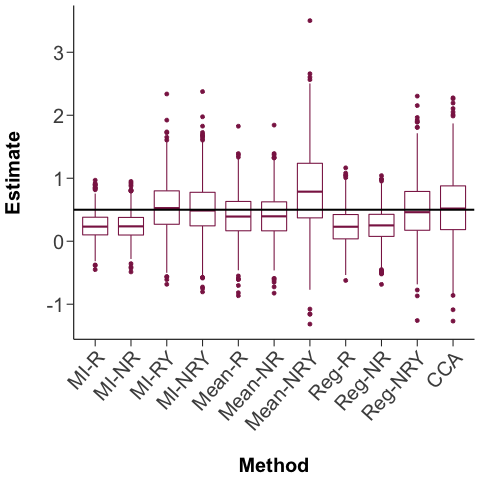}} 
    \vskip\baselineskip
    \subfigure{\includegraphics[width=0.46\textwidth]{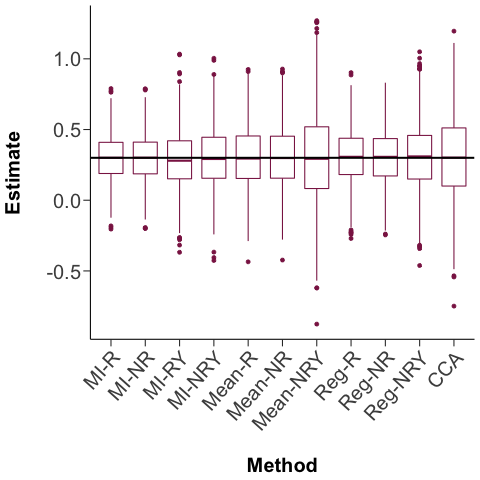}}
    \hspace{0.25in}
    \subfigure{\includegraphics[width=0.46\textwidth]{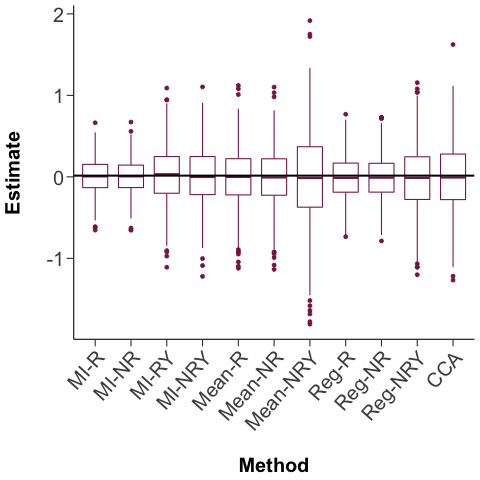}}
    \vspace{0.4\baselineskip}
    \caption[]{Results for scenario 3 under MAR. The top panel represents the distribution of $\beta_t$ (left) and $\beta_{tx_2}$ (right) estimates for the high association setting, where the true values are $\beta_t = 0.3$ and  $\beta_{tx_2} = 0.5$. The bottom panel represents the distribution of $\beta_t$ (left) and $\beta_{tx_2}$ (right) estimates for the low association setting, where the true values are $\beta_t = 0.3$ and  $\beta_{tx_2} = 0.015$.\label{fig:S3}}
\end{figure}

The top panels of Figure \ref{fig:S1} and Table \ref{tab:S1} summarize estimates under missingness scenario 1, when the covariate-outcome association is high. The bottom panels of Figure \ref{fig:S1} and Table \ref{tab:S1} summarize estimates under the same scenario, when the covariate-outcome association is low. Figure \ref{fig:S2} and Table \ref{tab:S2} summarize results under missingness scenario 2. Figure \ref{fig:S3} and Table \ref{tab:S3} do so for missingness scenario 3. In the high association setting, the averages of the $\beta_t$ and $\beta_{tx_2}$ estimates before deletion of data are 0.309 and 0.512, respectively. The MC-SD and 95\% CI coverage rates for the $\beta_t$ estimates before deletion are 0.198 and 0.958, respectively. The MC-SD and 95\% CI coverage rate for the $\beta_{tx_2}$ estimates before deletion are 0.327 and 0.951, respectively. In the low association setting, the averages of the $\beta_t$ and $\beta_{tx_2}$ estimates before deletion of data are 0.300 and 0.007, respectively. The MC-SD and 95\% CI coverage rates for the $\beta_t$ estimates before deletion are 0.193 and 0.940, respectively. The MC-SD and 95\% CI coverage rate for the $\beta_{tx_2}$ estimates before deletion are 0.263 and 0.953, respectively.

\FloatBarrier
\FloatBarrier
\FloatBarrier
\newpage

\FloatBarrier
\begin{table}[t]
\centering
\caption[]{Absolute biases, MC standard deviations, estimated standard errors, coverage probabilities, and average confidence lengths for $\beta_t$ and $\beta_{tx_2}$ estimates in scenario 1 under MAR.\label{tab:S1}}
\resizebox{\textwidth}{!}{
{\LARGE
\begin{tabular}{lcccccccccccc}
\hline
\multicolumn{1}{c}{} & \multicolumn{1}{c}{} & \multicolumn{5}{c}{$\beta_t$} & \multicolumn{1}{c}{} & \multicolumn{5}{c}{$\beta_{tx_2}$} \\ \cline{1-13}  
\multicolumn{2}{l}{\multirow{2}{*}{Method}} &  \multirow{2}{*}{\makecell{Absolute \\ bias}} & \multirow{2}{*}{MC-SD}      & \multirow{2}{*}{SE} & \multirow{2}{*}{Coverage}  & \multirow{2}{*}{\makecell{Average CI\\ length}}& \multirow{2}{*}{}     &  \multirow{2}{*}{\makecell{Absolute \\ bias}} & \multirow{2}{*}{MC-SD}      & \multirow{2}{*}{SE} & \multirow{2}{*}{Coverage}  & \multirow{2}{*}{\makecell{Average CI \\ length}}\\\\
 \cline{1-13}  \\ 
\multicolumn{13}{c}{High Association} \\
\cline{1-13}\\
MI-R                            &&0.039                                             & 0.182                  & 0.218                   & 0.984    &0.853                  &&0.212                                          & 0.239                     & 0.373                      & 0.984 &1.461                     \\
MI-NR                           &&0.039                                            & 0.182                      & 0.218                      & 0.981  &0.855                    &&0.212                                           & 0.240                   & 0.373                      & 0.987   & 1.461                \\
MI-RY                           &&0.006                                           & 0.221                   & 0.220                     & 0.950    &0.860                  &&0.015                                          & 0.441                     & 0.434                      & 0.954 &1.695                    \\ 
MI-NRY &&0.009 & 0.223 &0.223 &0.953 &0.872 &&0.013 &0.441 &0.435 &0.954 &1.704\\
Mean-R &&0.030 &0.235 &0.241 &0.957 &0.948 &&0.031 &0.394 &0.409 &0.962 & 1.606\\
Mean-NR &&0.030 &0.237 &0.247&0.958 &0.948 &&0.031  &0.397 &0.410 &0.961 &1.607\\
Mean-NRY &&0.086 &0.300 &0.236 &0.855&0.925 &&0.336 &0.667 &0.444 & 0.747 &1.739\\
Reg-R &&0.045 &0.208 &0.208 &0.948&0.815 &&0.213 &0.309 &0.313 &0.897 &1.229\\
Reg-NR &&0.042  &0.218 &0.218 &0.943&0.817 &&0.208 &0.312 &0.314 &0.908&1.232\\
Reg-NRY &&0.002 &0.246 &0.204 &0.889&0.800 &&0.014 &0.478 &0.335 &0.831&1.314\\
CCA    & \multicolumn{1}{c}{} &0.010 & \multicolumn{1}{c}{0.325} & \multicolumn{1}{c}{0.328} & \multicolumn{1}{c}{0.954} &  1.288 && 0.035 &\multicolumn{1}{c}{0.524} & \multicolumn{1}{c}{0.545} & \multicolumn{1}{c}{0.963}&2.145 \\\\ \hline \\
\multicolumn{13}{c}{Low Association} \\
\cline{1-13} \\
MI-R                            &&0.012                                             & 0.166                   & 0.197                   & 0.985                     & 0.774                                         && 0.005 &0.202                      & 0.303                     & 0.996&1.190                      \\
MI-NR                           &&0.011         & 0.167                     & 0.198                     & 0.984 &0.775                     &                                           &0.005& 0.203                     & 0.304                    & 0.997   &1.193                 \\
MI-RY                           &&0.015                                          & 0.214                      &0.208                     & 0.942                      &0.814                                           &&0.018 &0.338                     & 0.332                     & 0.947&1.295                     \\ 
MI-NRY && 0.017 &0.216 &0.209 &0.950&0.821&&0.019&0.343&0.334&0.947&1.311\\
Mean-R &&0.002 &0.220 &0.222 & 0.956&0.869 &&0.005&0.329 &0.332 &0.954&1.304\\
Mean-NR &&0.002 &0.222 &0.223 & 0.955&0.870 &&0.004& 0.329 &0.333 &0.954&1.304 \\
Mean-NRY &&0.004&0.325  &0.223 &0.855&0.872 &&0.002&0.550 & 0.333 &0.761&1.308\\
Reg-R &&0.002&0.190 &0.190 &0.953&0.744 &&0.003&0.255 &0.257 &0.950&1.009\\
Reg-NR &&0.002&0.193 &0.194 &0.951&0.745 &&0.004&0.256 &0.258 &0.955&1.011\\
Reg-NRY &&0.002&0.245 &0.190 &0.867&0.746 &&0.009&0.389 &0.258 &0.810&1.013\\
CCA    &&0.007&  \multicolumn{1}{c}{0.302} & \multicolumn{1}{c}{0.306} & \multicolumn{1}{c}{0.950} &1.202&&0.017& \multicolumn{1}{c}{0.403} & \multicolumn{1}{c}{0.415} & \multicolumn{1}{c}{0.958}&1.629 \\\\ \hline \\
\end{tabular}
}
}
\end{table}

\FloatBarrier
\begin{table}[t]
\centering
\caption[]{Absolute biases, MC standard deviations, estimated standard errors, coverage probabilities, and average confidence lengths for $\beta_t$ and $\beta_{tx_2}$ estimates in scenario 2 under MAR.\label{tab:S2}}
\resizebox{\textwidth}{!}{
{\LARGE
\begin{tabular}{lcccccccccccc}
\hline
\multicolumn{1}{c}{} & \multicolumn{1}{c}{} & \multicolumn{5}{c}{$\beta_t$} & \multicolumn{1}{c}{} & \multicolumn{5}{c}{$\beta_{tx_2}$} \\ \cline{1-13}  
\multicolumn{2}{l}{\multirow{2}{*}{Method}} &  \multirow{2}{*}{\makecell{Absolute \\ bias}} & \multirow{2}{*}{MC-SD}      & \multirow{2}{*}{SE} & \multirow{2}{*}{Coverage}  & \multirow{2}{*}{\makecell{Average CI\\ length}}& \multirow{2}{*}{}     &  \multirow{2}{*}{\makecell{Absolute \\ bias}} & \multirow{2}{*}{MC-SD}      & \multirow{2}{*}{SE} & \multirow{2}{*}{Coverage}  & \multirow{2}{*}{\makecell{Average CI \\ length}}\\\\
 \cline{1-13}  \\ 
\multicolumn{13}{c}{High Association} \\
\cline{1-13}\\
MI-R                            &&0.153                                             & 0.188                    & 0.210                   & 0.924&0.824                    &&0.470                                          & 0.259                     & 0.357                      & 0.805&1.399                     \\
MI-NR                           &&0.153                                            & 0.188                     & 0.211                      & 0.920 &0.827                     &&0.469                                           & 0.257                   & 0.358                     & 0.812&1.402                    \\
MI-RY                           &&0.001                                             & 0.213                    & 0.209                     & 0.949 &0.818                    &&0.019                                           & 0.412                    & 0.404                     & 0.940&1.576                      \\ 
MI-NRY &&0.003&0.213 &0.211&0.950&0.825 &&0.018&0.413&0.402&0.944&1.573\\
Mean-R&&0.057&0.229&0.227&0.931&0.892&&0.029&0.368 &0.380 &0.954&1.492\\
Mean-NR &&0.058&0.239 &0.238 &0.930&0.892 &&0.029&0.369 &0.382 &0.953&1.492 \\
Mean-NRY &&0.352& 0.262 &0.221 &0.619&0.869 &&0.835&0.578&0.409 &0.461&1.602\\
Reg-R &&0.184& 0.212 &0.207 &0.844&0.811 &&0.511&0.314 &0.315 &0.633&1.238\\
Reg-NR &&0.183&0.214 &0.217 &0.860&0.812 &&0.507&0.315 &0.316 &0.650&1.241\\
Reg-NRY &&0.011& 0.238 &0.203 &0.907&0.798&&0.004&0.452 &0.334 &0.861&1.310\\
CCA    && 0.007 &\multicolumn{1}{c}{0.294} & \multicolumn{1}{c}{0.283} & \multicolumn{1}{c}{0.944} & 1.112&&0.049&  \multicolumn{1}{c}{0.474} & \multicolumn{1}{c}{0.482} & \multicolumn{1}{c}{0.955}&1.897 \\\\ \hline \\
\multicolumn{13}{c}{Low Association} \\
\cline{1-13}  \\
MI-R                            &&0.005                                            & 0.173                    & 0.191                  & 0.967&0.749                     &&0.024                                          & 0.224                     & 0.287                      & 0.986&1.124                      \\
MI-NR                           &&0.005                                           & 0.173                     & 0.191                     & 0.967&0.750                      &&0.024                                           & 0.225                     & 0.287                    & 0.985&1.126                   \\
MI-RY                           &&0.003                                             & 0.196                     & 0.197                      & 0.954 &0.771                     &&0.003                                           & 0.306                     & 0.303                      & 0.941&1.185                     \\ 
MI-NRY &&0.002& 0.196 &0.198 &0.958&0.775 &&0.003&0.308 &0.305 &0.944&1.195\\
Mean-R &&0.001&0.206 &0.209 &0.954&0.819 &&0.007& 0.302 & 0.304 &0.945&1.191 \\
Mean-NR &&0.001& 0.208 &0.210 &0.955&0.820 &&0.006& 0.305 &0.309 &0.944&1.191\\
Mean-NRY &&0.018&0.274 &0.209 &0.859&0.821 &&0.029&0.447 &0.304 &0.824&1.193 \\
Reg-R &&0.006& 0.188 & 0.190 &0.949&0.744 &&0.016& 0.258 &0.257 &0.945&1.008\\
Reg-NR &&0.012& 0.192 &0.195 &0.955&0.744 &&0.025& 0.262 &0.265 &0.941&1.009 \\
Reg-NRY &&0.000 &0.224 & 0.190 &0.902&0.745 && 0.000&0.341 &0.258 &0.858&1.011\\
CCA  &&0.004 & \multicolumn{1}{c}{0.269} & \multicolumn{1}{c}{0.263} & \multicolumn{1}{c}{0.951} & 1.033 &&0.002&  \multicolumn{1}{c}{0.358} & \multicolumn{1}{c}{0.357} & \multicolumn{1}{c}{0.947}&1.401 \\\\ \hline \\
\end{tabular}
}
}
\end{table}

\FloatBarrier
\begin{table}[ht]
\centering
\caption[]{Absolute biases, MC standard deviations, estimated standard errors, coverage probabilities, and average confidence lengths for $\beta_t$ and $\beta_{tx_2}$ estimates in scenario 3 under MAR.\label{tab:S3}}
\resizebox{\textwidth}{!}{
{\LARGE
\begin{tabular}{lcccccccccccc}
\hline
\multicolumn{1}{c}{} & \multicolumn{1}{c}{} & \multicolumn{5}{c}{$\beta_t$} & \multicolumn{1}{c}{} & \multicolumn{5}{c}{$\beta_{tx_2}$} \\ \cline{1-13}  
\multicolumn{2}{l}{\multirow{2}{*}{Method}} &  \multirow{2}{*}{\makecell{Absolute \\ bias}} & \multirow{2}{*}{MC-SD}      & \multirow{2}{*}{SE} & \multirow{2}{*}{Coverage}  & \multirow{2}{*}{\makecell{Average CI\\ length}}& \multirow{2}{*}{}     &  \multirow{2}{*}{\makecell{Absolute \\ bias}} & \multirow{2}{*}{MC-SD}      & \multirow{2}{*}{SE} & \multirow{2}{*}{Coverage}  & \multirow{2}{*}{\makecell{Average CI \\ length}}\\\\
 \cline{1-13}  \\ 
\multicolumn{13}{c}{High Association} \\
\cline{1-13}\\
MI-R && 0.057 & 0.172  & 0.206  & 0.977 &0.806  && 0.259 & 0.214 &0.342                     & 0.966  &1.342                   \\
MI-NR   && 0.057 &0.173 &0.206 &0.978  &0.808  && 0.259 &0.213  &0.343 &0.972 &1.346                    \\
MI-RY   &&0.013  &0.219 & 0.212 & 0.942 &0.828  &&0.035 &0.406  & 0.402 & 0.947  &1.571                   \\ 
MI-NRY &&0.000 &0.222 & 0.213 &0.944 &0.835 &&0.008 &0.421 &0.411 &0.942 &1.611\\
Mean-R &&0.004 &0.230 &0.231 &0.951 &0.904 &&0.097 &0.358 &0.375 &0.949 &1.471\\
Mean-NR &&0.004 &0.230 &0.231 &0.951 &0.904 &&0.097 &0.359 &0.378 &0.954 &1.472\\
Mean-NRY &&0.080 &0.305 &0.227 &0.838 &0.891 &&0.310 &0.636 &0.407&0.750 &1.596\\
Reg-R &&0.070 &0.197 & 0.197 &0.927&0.772 &&0.266 &0.291 & 0.286 &0.849 &1.122\\
Reg-NR &&0.060 &0.202 &0.205 &0.938 &0.773 &&0.248 &0.294 &0.294 &0.880 &1.124\\
Reg-NRY &&0.012 &0.256 &0.194 &0.862 & 0.762 &&0.013 &0.478 &0.304 &0.785 &1.193\\
CCA    & \multicolumn{1}{c}{} & \multicolumn{1}{c}{0.010}& \multicolumn{1}{c}{0.325} & \multicolumn{1}{c}{0.328} & \multicolumn{1}{c}{0.954} & \multicolumn{1}{c}{1.288}& \multicolumn{1}{c}{} &  \multicolumn{1}{c}{0.035}&  \multicolumn{1}{c}{0.524} & \multicolumn{1}{c}{0.545} & \multicolumn{1}{c}{0.963}& \multicolumn{1}{c}{2.144} \\\\ \hline \\
\multicolumn{13}{c}{Low Association} \\
\cline{1-13} \\
MI-R &&0.001 & 0.164                    & 0.197                  & 0.981  &0.773                      &                                          &0.010 &0.203                      & 0.303                     & 0.996 &1.187                    \\
MI-NR                           &&0.001                                            & 0.164                    & 0.197                     & 0.981   &0.774                  && 0.010                                           & 0.202                    & 0.303                    & 0.996  &1.189                   \\
MI-RY                           &&0.016                                             & 0.211                   & 0.208                      & 0.940    &0.810                &&0.013                                           & 0.335                      & 0.329                    & 0.945    &1.284                 \\ 
MI-NRY &&0.001 &0.213 &0.209 &0.949 &0.820 &&0.007 &0.336 &0.333 &0.952 &1.308 \\
Mean-R &&0.002 &0.221 & 0.222 &0.946 &0.870 &&0.020 &0.329 & 0.332 &0.954 &1.303\\
Mean-NR &&0.003 &0.223 &0.226 &0.944 &0.871 &&0.021 &0.332 &0.337 &0.957 &1.303\\
Mean-NRY &&0.005 &0.325 &0.223 &0.823 &0.874 &&0.023 &0.548 &0.433 &0.764 &1.307\\
Reg-R &&0.006 &0.194 & 0.190 &0.942 &0.745 &&0.027 &0.256 &0.257 &0.941 &1.008\\
Reg-NR &&0.005 &0.197 & 0.195 &0.953 &0.746 &&0.024 &0.257 &0.258 &0.947 &1.010\\
Reg-NRY &&0.006 &0.242 &0.191 &0.877 &0.748 &&0.024 &0.386 &0.258 & 0.818 &1.012\\
\multicolumn{1}{l}{CCA}    & \multicolumn{1}{c}{} & \multicolumn{1}{c}{0.007}& \multicolumn{1}{c}{0.302} & \multicolumn{1}{c}{0.306} & \multicolumn{1}{c}{0.950} & \multicolumn{1}{c}{1.202}& \multicolumn{1}{c}{} &  \multicolumn{1}{c}{0.017}&  \multicolumn{1}{c}{0.403} & \multicolumn{1}{c}{0.415} & \multicolumn{1}{c}{0.958}& \multicolumn{1}{c}{1.629} \\\\ \hline \\
\end{tabular}
}
}
\end{table}

\section{Simulation results under the ICIN assumption} \label{sec:S3}
\begin{figure}[t]
    \centering
    \subfigure{\includegraphics[width=0.46\textwidth]{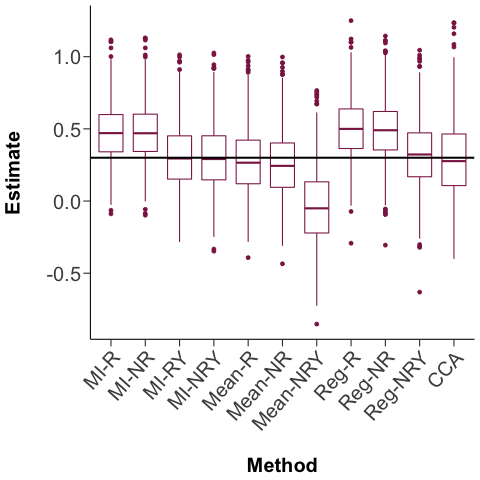}} 
    \hspace{0.25in}
    \subfigure{\includegraphics[width=0.46\textwidth]{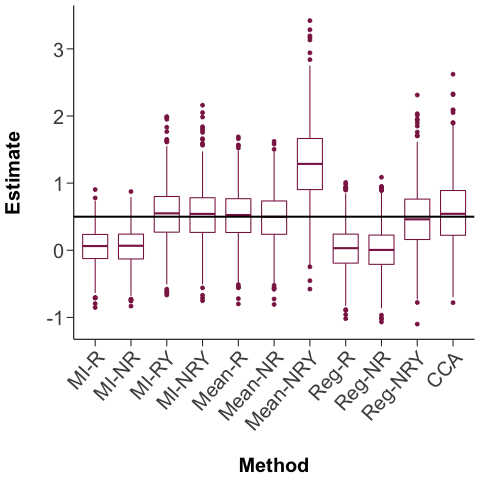}} 
    \vskip\baselineskip
    \subfigure{\includegraphics[width=0.46\textwidth]{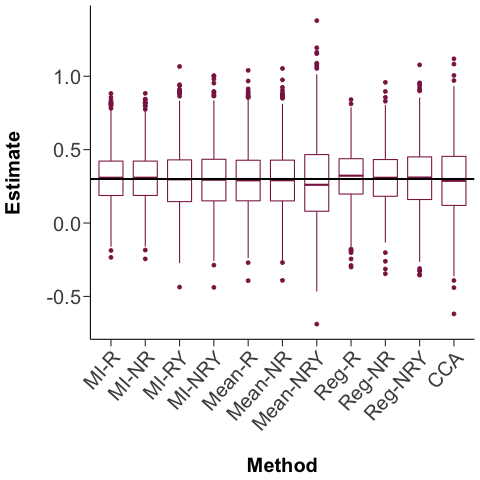}}
    \hspace{0.25in}
    \subfigure{\includegraphics[width=0.46\textwidth]{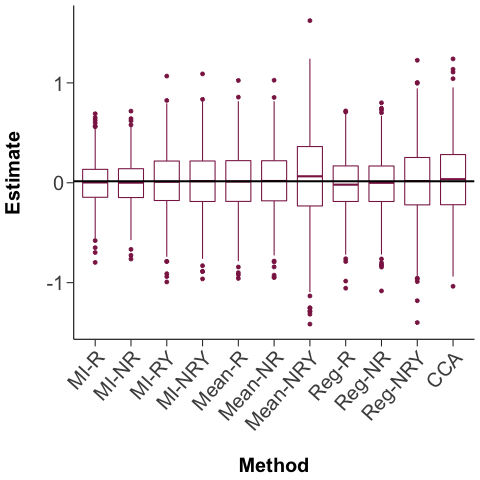}}
    \vspace{0.4\baselineskip}
    \caption[]{Results for scenario 2 under ICIN. The top panel represents the distribution of $\beta_t$ (left) and $\beta_{tx_2}$ (right) estimates for the high association setting, where the true values are $\beta_t = 0.3$ and  $\beta_{tx_2} = 0.5$. The bottom panel represents the distribution of $\beta_t$ (left) and $\beta_{tx_2}$ (right) estimates for the low association setting, where the true values are $\beta_t = 0.3$ and  $\beta_{tx_2} = 0.015$.\label{fig:S4}}
\end{figure}

\begin{figure}[t]
    \centering
    \subfigure{\includegraphics[width=0.46\textwidth]{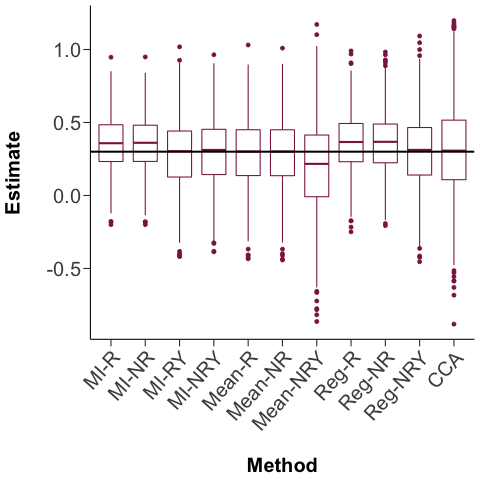}} 
    \hspace{0.25in}
    \subfigure{\includegraphics[width=0.46\textwidth]{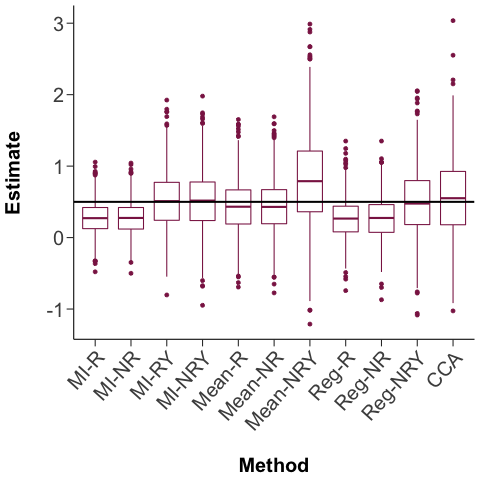}} 
    \vskip\baselineskip
    \subfigure{\includegraphics[width=0.46\textwidth]{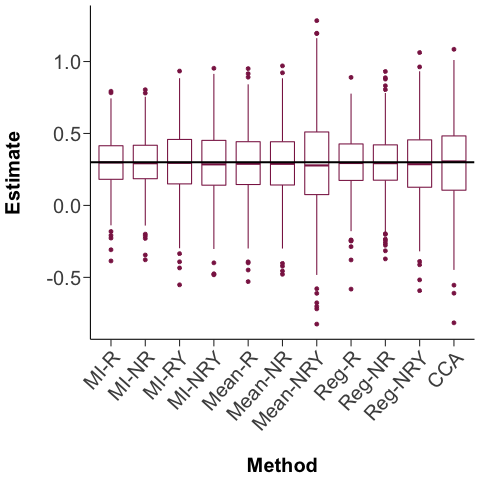}}
    \hspace{0.25in}
    \subfigure{\includegraphics[width=0.46\textwidth]{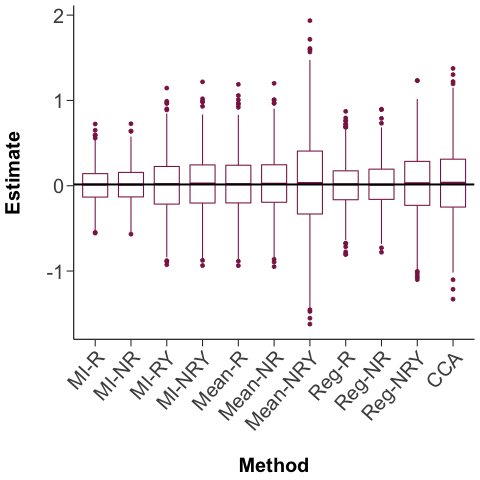}}
    \vspace{0.4\baselineskip}
    \caption[]{Results for scenario 3 under ICIN. The top panel represents the distribution of $\beta_t$ (left) and $\beta_{tx_2}$ (right) estimates for the high association setting, where the true values are $\beta_t = 0.3$ and  $\beta_{tx_2} = 0.5$. The bottom panel represents the distribution of $\beta_t$ (left) and $\beta_{tx_2}$ (right) estimates for the low association setting, where the true values are $\beta_t = 0.3$ and  $\beta_{tx_2} = 0.015$.\label{fig:S5}}
\end{figure}

Here, we provide graphical and tabular summaries for results under missingness scenarios 2 and 3. Figure \ref{fig:S4} and Table \ref{tab:S4} summarize estimates under missingness scenario 2. Figure \ref{fig:S5} and Table \ref{tab:S5} do so for missingness scenario 3.


\FloatBarrier
\begin{table}[t]
\centering
\caption[]{Absolute biases, MC standard deviations, estimated standard errors, coverage probabilities, and average confidence lengths for $\beta_t$ and $\beta_{tx_2}$ estimates in scenario 2 under ICIN.\label{tab:S4}}
\resizebox{\textwidth}{!}{
{\LARGE
\begin{tabular}{lcccccccccccc}
\hline
\multicolumn{1}{c}{} & \multicolumn{1}{c}{} & \multicolumn{5}{c}{$\beta_t$} & \multicolumn{1}{c}{} & \multicolumn{5}{c}{$\beta_{tx_2}$} \\ \cline{1-13}  
\multicolumn{2}{l}{\multirow{2}{*}{Method}} &  \multirow{2}{*}{\makecell{Absolute \\ bias}} & \multirow{2}{*}{MC-SD}      & \multirow{2}{*}{SE} & \multirow{2}{*}{Coverage}  & \multirow{2}{*}{\makecell{Average CI\\ length}}& \multirow{2}{*}{}     &  \multirow{2}{*}{\makecell{Absolute \\ bias}} & \multirow{2}{*}{MC-SD}      & \multirow{2}{*}{SE} & \multirow{2}{*}{Coverage}  & \multirow{2}{*}{\makecell{Average CI \\ length}}\\\\
 \cline{1-13}  \\ 
\multicolumn{13}{c}{High Association} \\
\cline{1-13}\\

MI-R &&0.170 &0.193 &0.211  &0.904 &0.828 &&0.442 &0.272  &0.357   &0.830 &1.399 \\
MI-NR  &&0.170  &0.194  &0.212  &0.902 &0.828 &&0.441 &0.274  &0.358  &0.825 &1.402 \\
MI-RY  &&0.003  &0.219  &0.211  &0.942  &0.825 &&0.017 &0.409 &0.401  &0.942 &1.565 \\ 
MI-NRY &&0.001  &0.221 &0.212 &0.940 &0.830 &&0.017  &0.416 &0.402 &0.952 &1.572\\
Mean-R &&0.029 & 0.226 &0.218 &0.943 &0.856 &&0.019 &0.385 &0.385 &0.952 &1.510\\
Mean-NR &&0.047&0.237 & 0.229 &0.940 &0.861 &&0.008 &0.389 &0.389 &0.951 &1.494\\
Mean-NRY &&0.336 &0.263 &0.215&0.614 &0.840 &&0.796 &0.581 &0.407 &0.483 &1.594\\
Reg-R &&0.205  &0.206 & 0.199 &0.815 &0.779 &&0.479 &0.330 &0.318 &0.674 &1.249\\
Reg-NR &&0.190  &0.209 &0.204 &0.828 &0.786 &&0.496 &0.337 &0.319 &0.652 &1.246\\
Reg-NRY &&0.020 &0.229 &0.199 & 0.907 &0.771 &&0.026 &0.471 &0.336 &0.832 &1.317\\
CCA    & \multicolumn{1}{c}{} & \multicolumn{1}{c}{0.007} & \multicolumn{1}{c}{0.266} & \multicolumn{1}{c}{0.256} & \multicolumn{1}{c}{0.944} &\multicolumn{1}{c}{1.005 }& \multicolumn{1}{c}{} &  \multicolumn{1}{c}{0.056 } &\multicolumn{1}{c}{0.495} & \multicolumn{1}{c}{0.495} & \multicolumn{1}{c}{0.954} & \multicolumn{1}{c}{1.948 }\\\\ \hline \\
\multicolumn{13}{c}{Low Association} \\
\cline{1-13} \\
MI-R    &&0.008       & 0.178       &0.196      & 0.970   &0.768    && 0.018   &0.219    &0.286                  & 0.991  &1.122 \\
MI-NR    &&0.007     &0.178      &0.196       &0.968  &0.769   &&0.017 &0.219  & 0.287                     & 0.992 &1.124 \\
MI-RY    & &0.002 & 0.209     & 0.202    & 0.944    &0.790 & &0.004 &0.297  & 0.301                     & 0.954   &1.176     \\ 
MI-NRY &&0.003  &0.210 &0.204 &0.944 &0.798 &&0.002 &0.297 & 0.304 &0.948 &1.192\\
Mean-R &&0.004  &0.201 &0.200 & 0.945 &0.784 &&0.005 &0.300 &0.303 &0.952 &1.189\\
Mean-NR &&0.005 &0.209 &0.202 &0.943 &0.792 &&0.005 & 0.305 &0.305 &0.952 &1.190 \\
Mean-NRY &&0.026 &0.277 & 0.202 & 0.860 &0.793 &&0.047 &0.449 &0.304 &0.804 &1.192 \\
Reg-R &&0.020 &0.181 & 0.182 & 0.951 &0.715 &&0.029 &0.262 & 0.256 & 0.953 &1.006 \\
Reg-NR &&0.013 &0.183 & 0.184 & 0.944 &0.720 &&0.018 &0.263 &0.257 &0.942 &1.008 \\
Reg-NRY &&0.004 &0.222 &0.184 &0.894 &0.721 &&0.001 &0.352 & 0.257 &0.853 &1.009 \\
CCA    & \multicolumn{1}{c}{} & \multicolumn{1}{c}{0.009} & \multicolumn{1}{c}{0.245} & \multicolumn{1}{c}{0.240} & \multicolumn{1}{c}{0.952} & \multicolumn{1}{c}{0.942} & \multicolumn{1}{c}{} &  \multicolumn{1}{c}{0.013} &  \multicolumn{1}{c}{0.353} & \multicolumn{1}{c}{0.357} & \multicolumn{1}{c}{0.961} & \multicolumn{1}{c}{1.403}\\\\ \hline \\
\end{tabular}
}
}
\end{table}


\FloatBarrier
\begin{table}[t]
\centering
\centering
\caption[]{Absolute biases, MC standard deviations, estimated standard errors, coverage probabilities, and average confidence lengths for $\beta_t$ and $\beta_{tx_2}$ estimates in scenario 3 under ICIN.\label{tab:S5}}
\resizebox{\textwidth}{!}{
{\LARGE
\begin{tabular}{lcccccccccccc}
\hline
\multicolumn{1}{c}{} & \multicolumn{1}{c}{} & \multicolumn{5}{c}{$\beta_t$} & \multicolumn{1}{c}{} & \multicolumn{5}{c}{$\beta_{tx_2}$} \\ \cline{1-13}  
\multicolumn{2}{l}{\multirow{2}{*}{Method}} &  \multirow{2}{*}{\makecell{Absolute \\ bias}} & \multirow{2}{*}{MC-SD}      & \multirow{2}{*}{SE} & \multirow{2}{*}{Coverage}  & \multirow{2}{*}{\makecell{Average CI\\ length}}& \multirow{2}{*}{}     &  \multirow{2}{*}{\makecell{Absolute \\ bias}} & \multirow{2}{*}{MC-SD}      & \multirow{2}{*}{SE} & \multirow{2}{*}{Coverage}  & \multirow{2}{*}{\makecell{Average CI \\ length}}\\\\
 \cline{1-13}  \\ 
\multicolumn{13}{c}{High Association} \\
\cline{1-13}\\

MI-R                            & &0.056  &0.178                    & 0.217                   & 0.984        &0.849             & &0.227       &0.222                      & 0.339                      & 0.970   &1.329                   \\
MI-NR                           &&0.056                                            & 0.179                     & 0.217                     & 0.984        &0.851            & &0.227                                          & 0.224                    & 0.340                     & 0.970      &1.334              \\
MI-RY                           &&0.009                                             & 0.227                      & 0.224                    & 0.947  &0.876                    && 0.014                                           & 0.395                    & 0.391                    & 0.948        &1.528             \\ 
MI-NRY &&0.000 &0.226 &0.226 &0.951 &0.885 && 0.012  & 0.407 &0.408 &0.947 &1.599 \\
Mean-R &&0.008 & 0.230 &0.229 &0.949 &0.898  &&0.073 &0.361 &0.373 &0.945&1.462\\
Mean-NR &&0.007 &0.230 &0.229 &0.951 &0.898 &&0.071 &0.363 &0.373 &0.944 &1.463\\
Mean-NRY &&0.092 &0.317 &0.226 &0.812 &0.888 &&0.294 &0.636 &0.396 &0.747 &1.553\\
Reg-R &&0.062  &0.197 &0.197 &0.938 &0.774 &&0.235 &0.282 &0.286 &0.874 &1.122\\
Reg-NR &&0.060 &0.197 &0.197 &0.941 &0.774 &&0.228 &0.297 &0.287 &0.870 &1.125\\
Reg-NRY &&0.007 &0.249 &0.194 &0.872 &0.762 &&0.009 &0.479 &0.300 &0.785 &1.178\\
CCA    & \multicolumn{1}{c}{} & \multicolumn{1}{c}{0.006}  & \multicolumn{1}{c}{0.315} & \multicolumn{1}{c}{0.296} & \multicolumn{1}{c}{0.934} & \multicolumn{1}{c}{1.164}  & \multicolumn{1}{c}{} &  \multicolumn{1}{c}{0.062} &  \multicolumn{1}{c}{0.549} & \multicolumn{1}{c}{0.555} & \multicolumn{1}{c}{0.948} & \multicolumn{1}{c}{2.188}  \\\\ \hline \\
\multicolumn{13}{c}{Low Association} \\
\cline{1-13} \\
MI-R                            & &0.000                                            & 0.165                  & 0.208                   & 0.984       &0.814              && 0.007                                         & 0.203                      & 0.301                     & 0.998     &1.180                 \\
MI-NR                           & &0.002                                           & 0.168                      & 0.208                      & 0.983                      &0.817 &&0.006                                           & 0.207                     & 0.302                      & 0.997    &1.186                \\
MI-RY                           & &0.002                                            & 0.216                      & 0.219                     & 0.959                      &0.855                        &&0.006                   & 0.325                      & 0.325                     & 0.952        &1.272             \\
MI-NRY &&0.005 & 0.220 &0.222 &0.959 &0.871 &&0.003 & 0.335 &0.333 &0.948&1.309 \\
Mean-R &&0.007 &0.216 &0.219 &0.956 &0.859 &&0.004 &0.331 &0.332 &0.957  &1.301\\
Mean-NR &&0.008 &0.217 &0.219 &0.957 &0.860 &&0.006 &0.331 & 0.332 &0.958 &1.301\\
Mean-NRY &&0.012 &0.321 &0.220 &0.816 &0.863 &&0.015 &0.550 &0.333 &0.756 &1.305\\
Reg-R && 0.002 &0.187 &0.189 &0.961 &0.741 &&0.005 &0.257 &0.257 &0.935 &1.007\\
Reg-NR &&0.005 &0.190 &0.189 &0.948 &0.742&&0.000 &0.258 &0.257 &0.953 &1.009\\
Reg-NRY &&0.007 &0.240 & 0.190 &0.879 &0.743 &&0.007 &0.386 &0.258 &0.803 &1.011\\
CCA    & \multicolumn{1}{c}{} & \multicolumn{1}{c}{0.002} & \multicolumn{1}{c}{0.286} & \multicolumn{1}{c}{0.279} & \multicolumn{1}{c}{0.947} &  \multicolumn{1}{c}{1.094} &\multicolumn{1}{c}{} &  \multicolumn{1}{c}{0.019}&  \multicolumn{1}{c}{0.419} & \multicolumn{1}{c}{0.416} & \multicolumn{1}{c}{0.940}& \multicolumn{1}{c}{1.631} \\\\ \hline \\
\end{tabular}
}
}
\end{table}